\documentclass[aps,prd,twocolumn,nofootinbib,preprintnumbers,superscriptaddress,longbibliography,nobibnotes]{revtex4-2}
\usepackage{style}

\begin{document}

\title{Determining Spin-Dependent Light Dark Matter Rates from Neutron Scattering}

\author{Asher Berlin\,\orcidlink{0000-0002-1156-1482}}
\email{aberlin@fnal.gov}
\affiliation{Theoretical Physics Division, Fermi National Accelerator Laboratory, Batavia, IL 60510, USA}
\affiliation{Superconducting Quantum Materials and Systems Center (SQMS), Fermi National Accelerator Laboratory, Batavia, IL 60510, USA}
\author{Alexander J. Millar\,\orcidlink{0000-0003-3526-0526}}
\email{amillar@fnal.gov}
\affiliation{Theoretical Physics Division, Fermi National Accelerator Laboratory, Batavia, IL 60510, USA}
\affiliation{Superconducting Quantum Materials and Systems Center (SQMS), Fermi National Accelerator Laboratory, Batavia, IL 60510, USA}
\author{Tanner Trickle\,\orcidlink{0000-0003-1371-4988}}
\email{ttrickle@fnal.gov}
\affiliation{Theoretical Physics Division, Fermi National Accelerator Laboratory, Batavia, IL 60510, USA}
\author{Kevin Zhou\,\orcidlink{0000-0002-9810-3977}}
\email{kzhou7@berkeley.edu}
\affiliation{Berkeley Center for Theoretical Physics, University of California, Berkeley, CA 94720, USA}
\affiliation{Theoretical Physics Group, Lawrence Berkeley National Laboratory, Berkeley, CA 94720, USA}

\begin{abstract}
\noindent
The scattering and absorption rates of light dark matter with electron spin-dependent interactions depend on the target's spin response.
We show how this response is encoded by the target's dynamical magnetic susceptibility, which can be measured using neutron scattering.
We directly use existing neutron scattering data to compute the dark matter scattering rate in a candidate target material, finding close agreement with the previous first-principles calculation at MeV dark matter masses.
Complementary experiments and measurements can extend the reach of this technique to other dark matter models and masses, and identify promising target materials for future experiments.
\end{abstract}

\maketitle

\section{Introduction}
\label{sec:introduction}

The nature of dark matter (DM) remains among the most profound unsolved mysteries in physics. As nuclear recoil based direct detection experiments~\cite{LZ:2022lsv,XENON:2024wpa} approach neutrino-background-limited sensitivity~\cite{OHare:2021utq}, it is imperative to find other avenues to search for theoretically well-motivated DM candidates. 
One class of such candidates is ``light" DM, with sub-GeV mass, whose relic abundance can be set by a variety of cosmological production mechanisms~\cite{Antypas:2022asj,Zurek:2024qfm}. Detection of light DM requires a new generation of direct detection experiments with lower thresholds, operating at the boundary of condensed matter and particle physics~\cite{Kahn:2021ttr}.

A variety of ongoing and upcoming experiments probe light DM-electron interactions in semiconducting crystals with $\eV$-scale band gaps~\cite{CDEX:2019exx,CDEX:2022kcd,DAMIC-M:2023gxo,EDELWEISS:2020fxc,SENSEI:2023zdf,SuperCDMS:2019jxx,SuperCDMS:2020ymb,Oscura:2022vmi,DAMIC-M:2025luv}. Such targets have sensitivity to DM scattering and absorption with masses as low as $\sim \MeV$ and $\sim \eV$, respectively. In this case, calculations of the DM-induced excitation rates are more involved than for nuclear recoils, as they depend in detail on the target's electronic states. First-principles calculations have been performed for the simplest DM models~\cite{Essig:2015cda,Catena:2019gfa,Griffin:2021znd,Trickle:2022fwt,Dreyer:2023ovn}, and recently effective field theory methods have been applied to more general DM-electron interactions~\cite{Krnjaic:2024bdd,Liang:2024ecw,Catena:2022fnk,Catena:2021qsr}. 

Targets with $\meV$-scale excitations have also been explored~\cite{Hochberg:2017wce,Coskuner:2019odd,Geilhufe:2019ndy,Inzani:2020szg,Chen:2022pyd,SuperCDMS:2022kse}. Collective excitations such as phonons~\cite{Schutz:2016tid,Knapen:2016cue,Knapen:2017ekk,Griffin:2018bjn,Trickle:2019nya,Campbell-Deem:2019hdx,Cox:2019cod,Kurinsky:2019pgb,Griffin:2019mvc,Baym:2020uos,Griffin:2020lgd,Mitridate:2020kly,Knapen:2021bwg,Campbell-Deem:2022fqm,Taufertshofer:2023rgq} and magnons~\cite{Trickle:2019ovy,Chigusa:2020gfs,Mitridate:2020kly,Esposito:2022bnu,Catinari:2024ekq,Marocco:2025eqw} naturally have $\sim (1 - 100) \ \meV$ energies, and could thereby extend the search for DM-electron interactions down to keV-scale and meV-scale DM masses for scattering and absorption, respectively. Again, for the simplest DM models, the scattering~\cite{Trickle:2019ovy,Esposito:2022bnu,Marocco:2025eqw} and absorption~\cite{Mitridate:2020kly,Gelmini:2020xir,Catinari:2024ekq} rates have been computed using first principles, and effective field theory methods have been developed for general interactions~\cite{Trickle:2020oki,Mitridate:2023izi}. 

It has long been known that some DM-electron interaction rates can be directly related to measurable properties of the target if the dark interactions are sufficiently similar to Standard Model (SM) interactions. For example, since a dark photon couples to electrons in a manner analogous to a normal photon, the dark photon absorption rate in a material depends only on its photoelectric cross section~\cite{Pospelov:2008jk,An:2013yua,An:2014twa}, which itself can be written in terms of 
the material's electrical response functions~\cite{An:2013yua,An:2014twa,Hochberg:2016ajh,Hochberg:2016sqx,Mitridate:2021ctr}. More recently, the idea to reformulate the DM-electron interaction rates in terms of measurable target properties has been extended to DM-electron scattering~\cite{Hochberg:2021pkt,Knapen:2021run,Catena:2024rym,Essig:2024ebk}. 

For DM dominantly coupled to the electron density, the interaction rates can be expressed in terms of the dielectric function $\varepsilon(\qv, \omega)$, where $\qv$ and $\omega$ are the momentum and energy transfer to the target.\footnote{They can also be formulated in terms of, e.g., the conductivity, index of refraction, ``energy loss function," photon self-energy, or electric susceptibility. For an isotropic nonmagnetic material, these quantities all contain equivalent information.} This perspective is conceptually natural because it automatically accounts for arbitrary electronic states, including single or multiparticle states, and collective excitations. It is also practically useful, since it identifies a quantity that can be measured to experimentally calibrate the rate.

In this work, we consider DM coupled to the electron \textit{spin} density, and show that both scattering and absorption rates can be rewritten in terms of the magnetic susceptibility tensor $\x_{ij}(\qv, \omega)$, which quantifies the spin response to an external magnetic field. Since neutrons dominantly couple to electrons through a magnetic dipole-dipole interaction, this susceptibility can be directly probed by neutron scattering~\cite{Ball_1994,lovesey1984theory,squires1996introduction,boothroyd2020}, a well-developed method with a variety of dedicated facilities. Due to their mass, neutrons are a good kinematic match to light DM. Thus, existing measurements probe $\x_{ij}(\qv, \omega)$ at the values of $\qv$ and $\omega$ that are relevant for light DM scattering.  Note that this is unlike measurements of $\eps(\qv, \omega)$ using photons and electrons, which dominantly probe much lower $|\qv|$. 

This work is organized as follows. In Sec.~\ref{sec:rates_and_susceptibilities}, we derive the spin-dependent DM scattering and absorption rates in terms of the magnetic susceptibility. In \Sec{connection_to_neutron_scattering}, we introduce a ``data-driven" approach for the determination of spin-dependent DM scattering rates. Using data previously collected with the MAPS spectrometer~\cite{ewings2019upgrade} at the ISIS Neutron and Muon Source, we evaluate the scattering rate in \ce{Y3Fe5O12} (yttrium iron garnet, or YIG) for light DM with a dark magnetic dipole or anapole moment. The results closely agree with the rates originally computed in Ref.~\cite{Trickle:2019ovy}, which modeled the spin response with a Heisenberg Hamiltonian. Finally, in \Sec{discussion}, we discuss how these results can be generalized to other experiments, DM models, and target materials. Technical derivations are deferred to the appendices, which are referred to throughout the main text.

\section{Spin-Dependent Interactions And The Magnetic Susceptibility}
\label{sec:rates_and_susceptibilities}

Here we show how the magnetic susceptibility can be used to compute DM-electron interaction rates when the DM couples directly to electron spin. Such interactions can be described by an interaction Hamiltonian density
\begin{equation}
    \mathcal{H}_e(\mathbf{x}, t) = -\bm{\Phi}(\mathbf{x}, t) \cdot \mathbf{s}_e(\mathbf{x}, t) \, ,
    \label{eq:H_int_density}
\end{equation}
acting solely on the electrons. Here,
\be
\mathbf{s}_e(\mathbf{x}, t) \equiv \psi_e^\dagger(\mathbf{x}, t) \, \bm{\sigma} \, \psi_e(\mathbf{x}, t)
\ee
is twice the electron spin density, $\psi_e(\mathbf{x}, t)$ is the two-component electron field, $\bm{\sigma}$ are the Pauli matrices, and $\bm{\Phi}(\mathbf{x}, t)$ is a classical external potential. For example, the interaction between an electron and a background magnetic field $\mathbf{B}$ corresponds to $\bm{\Phi} = \mu_B \, \mathbf{B}$, where $\mu_B = |e| / (2 m_e)$ is the Bohr magneton.

DM-electron spin interactions can also be written as an interaction Hamiltonian in the form of \Eq{H_int_density}. For example, consider the Hamiltonian density which depends on both the DM fields and electron spin as,
\be
\label{eq:HintDM}
\mathcal{H}_{\x e} (\xv , t) = - \bm{X}_\x \cdot \bm{s}_e (\xv , t)
\, ,
\ee
where $\bm{X}_\x$ is an operator that depends on the DM field. To obtain the electron interaction Hamiltonian of \Eq{H_int_density}, we evaluate the matrix element for the process where DM transitions between the states $|\x_i\rangle \rightarrow | \x_f \rangle$, 
\be
\label{eq:HefromHxe}
\mathcal{H}_e = \frac{\langle \x_f | \mathcal{H}_{\x e} | \x_i \rangle}{ \sqrt{\langle \x_f|\x_f \rangle \,  \langle \x_i | \x_i \rangle}}
\, .
\ee
The corresponding interaction potential is thus obtained similarly from $\bm{X}_\x$ as
\begin{align}
\label{eq:PhiDef}
    \bm{\Phi}(\mathbf{x}, t) = \frac{\langle \x_f | \bm{X}_\x | \x_i \rangle}{ \sqrt{\langle \x_f|\x_f \rangle \,  \langle \x_i | \x_i \rangle}}
\, .
\end{align}
It will be convenient to decompose this potential as  
\begin{equation} \label{eq:dm_op_def}
\bm{\Phi}(\mathbf{x}, t) = \bm{\mathcal{O}}(\mathbf{q}, \omega_{\qv}) \, e^{-i q \cdot x}  \times\begin{cases} 1/V & \!\text{(scatter)} \\ 1/\sqrt{2 m_\x V} & \!\text{(absorb)} 
\end{cases}
\end{equation}
for a DM scattering or absorption interaction, respectively. Above, $q^\mu = (\omega_\qv, \qv)$ is the four-momentum deposited by the DM, $V$ is the target volume, $\bm{\mathcal{O}}$ depends on the DM model, and the overall factors are a convenient choice accounting for our normalization of the DM states in the denominator of \Eq{PhiDef}. Further details of this formalism are provided in App.~\ref{app:calcs}. 

The response of a target to a weak potential $\bm{\Phi}(\mathbf{x}, t)$ is quantified by linear response theory~\cite{coleman2015}. In Fourier space and component notation, the change in spin density is
\begin{align}
    \langle \, \delta s^i_e(\mathbf{q}, \omega) \, \rangle \approx \x_{ij}(\mathbf{q}, \omega) \, \Phi^j(\mathbf{q}, \omega) \, ,
    \label{eq:linear_response}
\end{align}
where $\x_{ij}(\mathbf{q}, \omega)$ is the target's (spin) magnetic susceptibility, $\delta \mathbf{s}_e(\mathbf{q}, \omega)$ is the Fourier transform of $\delta \mathbf{s}_e(\mathbf{x}, t) \equiv \mathbf{s}_e(\mathbf{x}, t) - \mathbf{s}_e(\bm{x}, 0)$, and the expectation value is over the target electronic states. Since $\x_{ij}$ encodes the target's spin response, it is no surprise that it determines the rate for any DM-electron interaction in the form of \Eq{H_int_density}. Analogous conclusions have been drawn when $\mathbf{s}_e(\mathbf{x}, t)$ is replaced in \Eq{H_int_density} with the electron number density, $n_e(\mathbf{x}, t) = \psi^\dagger_e(\mathbf{x}, t) \, \psi_e(\mathbf{x}, t)$. In that case, the DM-electron interactions are related to the electric susceptibility via the dielectric function~\cite{Knapen:2021run,Hochberg:2021pkt}. One can further define generalized susceptibilities which incorporate the response to other operators, as has been explored in Ref.~\cite{Catena:2024rym}. 

The DM-electron interaction rates will depend on the absorptive (anti-Hermitian) part of the magnetic susceptibility, $\x_{ij}''(\mathbf{q}, \omega) \equiv -i \left[ \x_{ij}(\mathbf{q}, \omega) - \x_{ji}^*(\mathbf{q}, \omega) \right] / 2$. As shown in App.~\ref{app:suscep}, it can be expressed in terms of the target's electronic states by
\begin{align}
    \x''_{ij}(\mathbf{q}, \omega) = \frac{\pi}{V} \sum_{ee'} ~ & \langle e | s^j_e(\mathbf{q}) | e' \rangle \, \langle e' | s^i_e(-\mathbf{q}) | e \rangle 
    \nonumber \\
    & \times  \, \delta(\omega - (E_{e'} - E_e)) \, ,
    \label{eq:chi_pp}
\end{align}
where $| e \rangle$ and $| e'\rangle$ are the initial and final electronic states, and $\v{s}_e(\v{q})$ is the Fourier transform of $\v{s}_e(\v{x}, 0)$. Here, we have implicitly assumed that the target is cold, so that DM can only \emph{deposit} energy and the sum only includes states with electronic energies $E_{e'} > E_e$. 

As a simple toy example, in a cubic Heisenberg ferromagnet with spin density $n_s$, magnetized along the $z$-axis and at zero temperature, the contribution to $\x''_{ij}(\mathbf{q}, \omega)$ due to single magnon production is
\begin{equation} \label{eq:one_magnon_1}
\x''_{xx} = \x''_{yy} = -i \x_{xy}'' = i \x_{yx}'' = 2 \pi \, n_s \, \delta(\w - \w_m(\qv)) \, ,
\end{equation}
where $\w_m(\qv)$ is the magnon frequency (related to $\qv$ via its dispersion relation), and we have assumed that losses are small. For further discussion of models of magnetic materials, see App.~\ref{app:spin_resp_model} and, e.g., Ref.~\cite{Marocco:2025eqw}.

\subsection{Scattering}
\label{subsec:suscep_scatter}

\renewcommand{\arraystretch}{2.6}
\setlength{\tabcolsep}{8pt}
\begin{table*}
\begin{center}
\begin{tabular}{@{}lll@{}} \toprule
Model & Lagrangian & form factor $\mathcal{F}^{ij}$ \\ \toprule
Magnetic dipole DM & $\dfrac{g_\x}{4 \, m_\x} \, \bar{\Psi}_\x \sigma^{\mu\nu} \Psi_\x \, F'_{\mu\nu} + g_e \, \bar{\Psi}_e \gamma^\mu \Psi_e A'_\mu$ & $2\left( \dfrac{g_\x g_e}{|\qv|^2 + m_{\text{med}}^2} \right)^2 \left( \dfrac{|\qv|^2}{4 \, m_\x m_e} \right)^2 (\delta^{ij} - \hat{\qv}^i \hat{\qv}^j)$ \\
Anapole DM & $\dfrac{g_\x}{4 \, m_\x^2} \, \bar{\Psi}_\x \gamma^\mu \gamma^5 \Psi_\x \, \del^\nu F'_{\mu\nu} + g_e \, \bar{\Psi}_e \gamma^\mu \Psi_e A'_\mu$ & $2\left( \dfrac{g_\x g_e}{|\qv|^2 + m_{\text{med}}^2} \right)^2 \left( \dfrac{|\qv|^3}{8 \, m_\x^2 m_e} \right)^2 (\delta^{ij} - \hat{\qv}^i \hat{\qv}^j)$ \\[2mm] \hline
Axial vector mediator $V_\mu$ & $V_\mu \, (g_\x \, \bar{\Psi}_\x \gamma^\mu \gamma^5 \Psi_\x + g_e \, \bar{\Psi}_e \gamma^\mu \gamma^5 \Psi_e)$ & $2\left( \dfrac{g_\x g_e}{|\qv|^2 + m_{\text{med}}^2} \right)^2 \delta^{ij}$ \\[2mm] \hline
Pseudoscalar mediator $\phi$ & $\phi \, (g_\x \, \bar{\Psi}_\x i \gamma^5 \Psi_\x + g_e \, \bar{\Psi}_e i \gamma^5 \Psi_e)$ & $2\left( \dfrac{g_\x g_e}{|\qv|^2 + m_{\text{med}}^2} \right)^2 \left( \dfrac{|\qv|^2}{4 \, m_\x m_e} \right)^2 \hat{\qv}^i \hat{\qv}^j$ \\
CP violating scalar mediator $\phi$ & $\phi \, (g_\x \, \bar{\Psi}_\x \Psi_\x + g_e \, \bar{\Psi}_e i \gamma^5 \Psi_e)$ & $2\left( \dfrac{g_\x g_e}{|\qv|^2 + m_{\text{med}}^2} \right)^2 \left(\dfrac{|\qv|}{2 \, m_e} \right)^2 \hat{\qv}^i \hat{\qv}^j$ \\
Dark electron EDM & $g_\x \, \bar{\Psi}_\x \gamma^\mu \Psi_\x A'_\mu + \dfrac{g_e}{4 \, m_e} \, i \bar{\Psi}_e \sigma^{\mu\nu} \gamma^5 \Psi_e \, F'_{\mu\nu}$ & $2\left( \dfrac{g_\x g_e}{|\qv|^2 + m_{\text{med}}^2} \right)^2 \left( \dfrac{|\qv|}{2 \, m_e} \right)^2 \hat{\qv}^i \hat{\qv}^j$ \\[1mm] \bottomrule 
\end{tabular}
\end{center}
\caption{Simple examples of models with spin-$1/2$ DM, $\x$, which yield a dominantly spin-dependent coupling to electrons in the nonrelativistic limit, organized by the tensor structure of their form factor $\mathcal{F}^{ij}$, as defined in \Eq{F_def}. In all cases, $m_{\text{med}}$ is the mediator mass, $m_\x$ is the DM mass, $\Psi_e$ is the electron Dirac field, $\Psi_\x$ is the DM Dirac field, and $\qv$ is the momentum transfer. For rows involving $A'_\mu$, which could be either the dark photon or the ordinary photon ($m_{\text{med}} \to 0$), $F'_{\mu\nu}$ is the field strength tensor, and $\sigma^{\mu\nu} = i [\gamma^\mu, \gamma^\nu]/2$. See Sec.~\ref{subsec:ns_derivation} for an example of how a form factor is computed and Sec.~\ref{sec:other_ints} for a discussion of the implications of the tensor structure of each form factor.}
\label{tab:models}
\end{table*}

We first consider DM scattering rates. Let the incoming DM particle $\x$ (not to be confused with the susceptibility tensor) have mass $m_\x$, nonrelativistic velocity $\mathbf{v}$, momentum $\mathbf{p} \approx m_\x \mathbf{v}$, kinetic energy $E \approx m_\x |\mathbf{v}|^2/2$, and spin state $s$. The DM deposits momentum $\mathbf{q}$ and energy 
\begin{equation} \label{eq:energy_deposit}
\omega_{\qv} \approx \mathbf{q} \cdot \mathbf{v} - \frac{|\mathbf{q}|^2}{2 m_\x}
\end{equation}
in the target, and exits with momentum $\mathbf{p}' = \mathbf{p} - \mathbf{q}$, kinetic energy $E' = E - \omega_{\qv}$, and spin state $s'$.

As shown in App.~\ref{app:rates}, the scattering rate $\Gamma_{\text{s}}(\mathbf{v})$ per incoming DM particle can be found by applying Fermi's golden rule to the interaction of \Eq{H_int_density}, yielding
\begin{align}
    \Gamma_{\text{s}}(\mathbf{v}) = \int \frac{\dd^3 \mathbf{q}}{(2 \pi)^3} ~ \mathcal{F}^{ij}(\mathbf{q}, \omega_{\qv}) \, \x''_{ij}(\mathbf{q}, \omega_{\qv}) \, ,
    \label{eq:scattering_rate}
\end{align}
where 
\begin{align}
\mathcal{F}^{ij}(\mathbf{q}, \omega_{\qv}) \equiv \frac{2}{2 S_\DM + 1} \sum_{ss'} \mathcal{O}^j(\mathbf{q}, \omega_{\qv}) \,\mathcal{O}^{* \, i}(\mathbf{q}, \omega_{\qv}) 
\, ,
    \label{eq:F_def}
\end{align}
is a form factor that depends on the DM-electron interaction, and $S_\DM$ is the DM spin. Here, the operator $\mathcal{O}^i$, defined in \Eq{dm_op_def}, can depend on $s$ and $s'$. In Table~\ref{tab:models}, we list some simple DM models which give rise to an interaction of the form \Eq{H_int_density}, along with their corresponding form factors. We outline a general procedure to compute $\mathcal{O}^i$ in App.~\ref{app:calcs_potential} and apply this to the case where DM interacts via a magnetic dipole in Sec.~\ref{subsec:ns_derivation}. 

The total velocity-averaged DM scattering rate per target mass is given by
\be
\label{eq:Rgen}
R_\text{s} = \frac{\rhodm}{m_\x \, \rhoT} \int \dd^3\vv ~ f_\DM(\vv) \, \Gamma_{\text{s}}(\vv) 
~,
\ee
where $\rhodm \approx 0.4 \ \GeV/\cm^3$ is the local DM density, $\rhoT$ is the mass density of the target, and $f_{\DM}(\vv)$ is the unit-normalized DM velocity distribution, which we take to be a boosted and truncated Maxwell--Boltzmann distribution,
\begin{align}
    f_{\DM}(\vv) \propto e^{- |\vv + \vv_E|^2 / v_0^2} \, \Theta(v_\text{esc} - |\vv + \vv_E|) \, .
    \label{eq:DM_velocity_distribution}
\end{align}
Here, $v_0 \approx 220 \, \text{km}/ \text{s}$ is the dispersion velocity, $|\mathbf{v}_E| \approx 240 \, \text{km} / \text{s}$ is the Earth velocity in the Galactic frame, and $v_\text{esc} \approx 500 \, \text{km} / \text{s}$ is the Galactic escape velocity~\cite{Trickle:2019ovy}.

\subsection{Absorption}
\label{subsec:suscep_absorb}

Although not the main focus of this work, let us now briefly consider the case where a nonrelativistic DM particle is absorbed and deposits its mass energy, corresponding to $\omega_{\qv} \approx m_\x$ and $\qv \approx m_\x \mathbf{v}$. In this case, applying Fermi's golden rule to \Eq{H_int_density} yields an absorption rate per incoming particle of
\begin{align} \label{eq:abs_rate}
    \Gamma_\text{a}(\mathbf{v}) = \frac{1}{2 m_\x} \, \mathcal{F}^{ij}(\mathbf{q}, m_\x) \, \x''_{ij}(\mathbf{q}, m_\x)\, .
\end{align}
The total absorption rate per unit target mass is then given by $R_\text{a} = (\rho_{\DM} / \rhoT) \Gamma_\text{a} / m_\x$.

As a quick example, consider the case where DM is an axion $a$ coupled to electrons via the ``axion wind" term. The relevant nonrelativistic interaction Hamiltonian density involving both the axion and electron fields, as in \Eq{HintDM}, is given by~\cite{Berlin:2023ubt}
\be
\label{eq:ae_HDM}
\mathcal{H}_{a e} \approx - g_{ae} \, \grad a \cdot \mathbf{s}_e 
\, ,
\ee
where $g_{ae}$ is a dimensionful coupling.\footnote{In \Eq{ae_HDM} we have neglected the ``axioelectric" term~\cite{Berlin:2023ubt}, whose absorption rate in simple non-magnetically ordered targets is related to the dielectric function~\cite{Pospelov:2008jk,An:2013yua,An:2014twa,Hochberg:2016ajh,Hochberg:2016sqx,Mitridate:2021ctr}.} To determine the Hamiltonian density $\mathcal{H}_{e}$ of \Eq{H_int_density}, which only depends on the electron degrees of freedom, we now evaluate the matrix element of $\mathcal{H}_{a e}$ with axion DM states $|a_i \rangle = |\qv \rangle$ and $|a_f \rangle = |0 \rangle$, as in \Eq{HefromHxe}. Using the standard relativistic state normalization $\langle \mathbf{q} | \mathbf{q} \rangle = 2 m_a V$~\cite{Peskin:1995ev} and the second line of \Eq{dm_op_def} then yields $\bm{\mathcal{O}} = i \, g_{ae} \, \qv$, where $m_a$ is the axion mass. Using this in \Eq{F_def} to determine the DM tensor $\mathcal{F}_{ij}$, the absorption rate of \Eq{abs_rate} reduces to
\begin{align}
\Gamma_\text{a}(\mathbf{v}) = g_{ae}^2 \, m_a \, v^i v^j \, \x_{ij}'' (\qv , m_a)  
\, .
\end{align}
For an infinite medium in the absorption kinematics limit, we have $\x''_{ij} \approx \mu_B^{-2} \, \text{Im}(-\mu_{ij}^{-1})$, where $\mu_{ij}$ is the magnetic permeability.\footnote{To see this, note that the classical dynamic spin magnetization density is $\v{M} = \mu_B \langle \delta \v{s}_e \rangle$. Therefore \Eq{linear_response} implies that $\v{M} = \mu_B^2 \chi \v{B}_{\text{eff}}$ in an effective magnetic field. Additionally, $\v{M} = (1 - \mu^{-1}) (\v{B}_{\text{ind}} + \v{B}_{\text{eff}})$ where $\v{B}_{\text{ind}}$ is the induced field. For a uniform medium in the absorption limit $|\qv| \to 0$, induced currents are confined to the medium's boundary. They can be neglected for an infinite medium, so that we can identify $\mu_B^2 \chi = 1 - \mu^{-1}$.} This recovers the result previously derived in Ref.~\cite{Berlin:2023ubt} using other methods. 

\section{Calibrating Dark Matter Scattering With Neutron Scattering}
\label{sec:connection_to_neutron_scattering}

In this section, we discuss how the spin-dependent DM scattering rate in YIG\footnote{YIG is a well-studied ferrimagnetic material known for its very low spin wave damping, which makes it an important model material in magnonics~\cite{cherepanov1993saga}.} can be inferred from neutron scattering measurements. We will use the raw data of Ref.~\cite{princep2017full}, taken on a single crystal sample of YIG with the MAPS spectrometer~\cite{ewings2019upgrade} at the ISIS Neutron and Muon Source. In Sec.~\ref{subsec:ns_derivation}, we derive the neutron scattering rate in terms of the form factor defined in \Eq{F_def}. We then relate it to the quantities reported in a neutron scattering experiment. In Sec.~\ref{subsec:connection_to_data}, we compare the kinematic coverage of the neutron scattering dataset to the region relevant for DM scattering. Though neutrons are a fairly good kinematic match for light DM, the dataset is incomplete in several ways, and we introduce extrapolation schemes to overcome this. In Sec.~\ref{subsec:sensitivity_proj}, we evaluate the DM scattering rate numerically using our data-driven approach, and compare it to the result of the first principles calculation of Ref.~\cite{Trickle:2019ovy}.

\subsection{The Neutron Scattering Rate}
\label{subsec:ns_derivation}

\subsubsection*{Derivation of the Form Factor}

Neutrons predominantly interact with electrons through their magnetic dipole moment. The corresponding neutron-electron interaction Lagrangian is the same as the ``magnetic dipole DM'' model in Table~\ref{tab:models}, upon replacing the dark matter with the neutron and the dark photon with the ordinary photon. This amounts to substituting in the first row of Table~\ref{tab:models}: $\Psi_\x \to \Psi_n$, $m_\x \to m_n$, $g_e \to e$ (where $e = - |e|$), $m_{\text{med}} \to 0$, and $g_\x \to \gamma_n e$ where $\gamma_n \approx -1.913$, so that
\begin{equation}
    \mathcal{L}_{ne} \supset \frac{\gamma_n e}{4 \, m_n} \, \bar{\Psi}_n \sigma^{\mu \nu} \Psi_n F_{\mu \nu} - e A_\mu \bar{\Psi}_e \gamma^\mu \Psi_e \, ,
    \label{eq:L_ne}
\end{equation}
where $\Psi_{\x, n, e}$ is the Dirac field for a DM particle, neutron, or electron, respectively. 
The resulting scattering rate is well-known (see Refs.~\cite{lovesey1984theory,squires1996introduction,boothroyd2020} for pedagogical discussions), but for illustration we will determine it by computing the corresponding form factor $\mathcal{F}^{ij}$.

In the low-energy limit, the dominant interaction arising from \Eq{L_ne} is the spin-spin coupling mediated by the magnetic field $\B$, corresponding to a Hamiltonian density
\begin{equation} \label{eq:H_ne}
    \mathcal{H}_{ne} = \left( \frac{\gamma_n e}{2 m_n} \, \psi_n^\dagger \bm{\sigma} \psi_n - \frac{e}{2 m_e} \, \psi_e^\dagger \bm{\sigma} \psi_e \right) \cdot \B \, ,
\end{equation}
where $\psi_n$ and $\psi_e$ are the two-component neutron and electron fields. An effective interaction Hamiltonian, which captures the tree-level scattering process of \Eq{H_ne}, is found by ``integrating out" the photon in the limit that $|\qv| \gg \omega_\qv$ (corresponding to the kinematics of nonrelativistic scattering),\footnote{More technically, this arises from matching to a diagram involving the exchange of a virtual photon. For a detailed derivation of a similar Hamiltonian, see App.~\ref{app:calcs}.}
\begin{align}
    \mathcal{H}_{ne} &\approx \frac{\gamma_n e^2}{4 m_n m_e} \, (\psi_n^\dagger (\qv \times \bm{\sigma})^i \psi_n) \, \frac{\delta^{ij}}{|\qv|^2} \, (\psi_e^\dagger (\qv \times \bm{\sigma})^j \psi_e) \nonumber \\
    &= \frac{\gamma_n e^2}{4 m_n m_e} \, P_T^{ij}\,  ( \psi_n^\dagger \sigma^i \psi_n ) \, ( \psi_e^\dagger \sigma^j \psi_e )  \, , \label{eq:H_ne_eff}
\end{align}
where we defined the transverse projector
\begin{equation}
P_T^{ij} = \delta^{ij} - \hat{\qv}^i \hat{\qv}^j
\, .
\end{equation}

Analogous to the above examples, we determine the Hamiltonian density $\mathcal{H}_{e}$ of \Eq{H_int_density} relevant for neutron scattering by evaluating the matrix element of $\mathcal{H}_{n e}$ with neutron states $|n_i \rangle = |\pv s \rangle$ and $|n_f \rangle = |\pv^\prime s^\prime \rangle$ (as in \Eq{HefromHxe} but with $| \x_{i,f} \rangle \to | n_{i,f} \rangle$). This gives $\mathcal{H}_e = \langle \mathbf{p}' s' | \mathcal{H}_{ne} | \mathbf{p} s \rangle / (2 m_n V)$, so that
\begin{equation}
\mathcal{O}^i = -\frac{\gamma_n e^2}{4 m_n m_e} \, P_T^{ij} \sigma^j_{ss'}
\, .
\end{equation}
Applying \Eq{F_def} yields the form factor $\mathcal{F}^{ij}$ listed in the first row of Table~\ref{tab:models}. Using \Eq{scattering_rate}, this corresponds to a neutron scattering rate
\begin{align}
    \Gamma_n(\vv) = \frac{\gamma_n^2 e^4}{8 \, m_n^2 m_e^2} \int \frac{\dd^3\qv}{(2 \pi)^3} ~ P_T^{ij} \x''_{ij}(\qv, \omega_{\qv}) \, ,
    \label{eq:neutron_scattering_rate_sus}
\end{align}
where $\omega_{\qv}$ is given by \Eq{energy_deposit} with $m_\x \to m_n$. 

Here we have focused on the neutron's interaction with the electron spin. In App.~\ref{app:subdominant_ints}, we discuss why other interactions are subdominant, and why this is also true for the DM models in Table~\ref{tab:models}.

\begin{figure*}[ht!]
    \centering
    \includegraphics[width=0.45\linewidth]{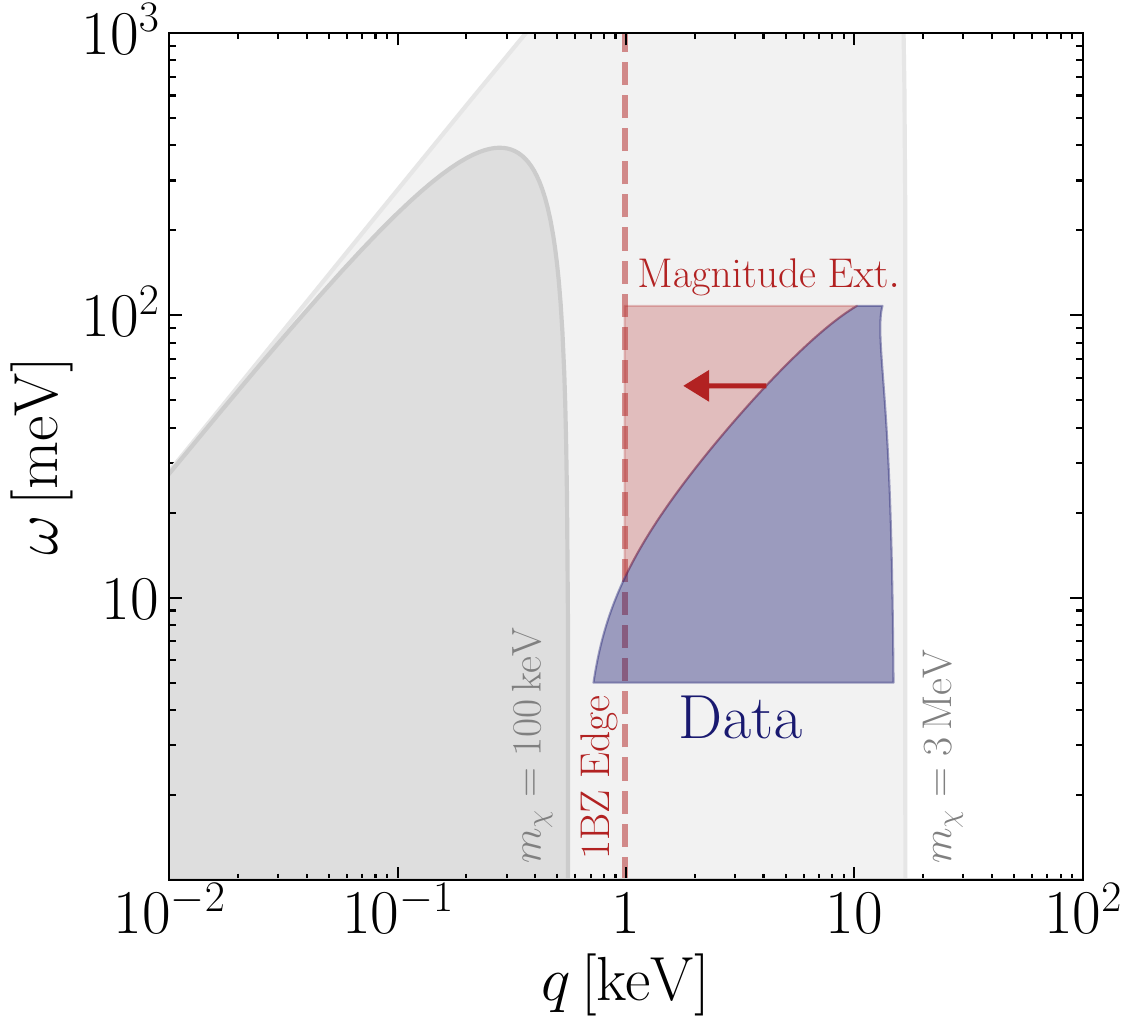}
    \includegraphics[width=0.45\linewidth]{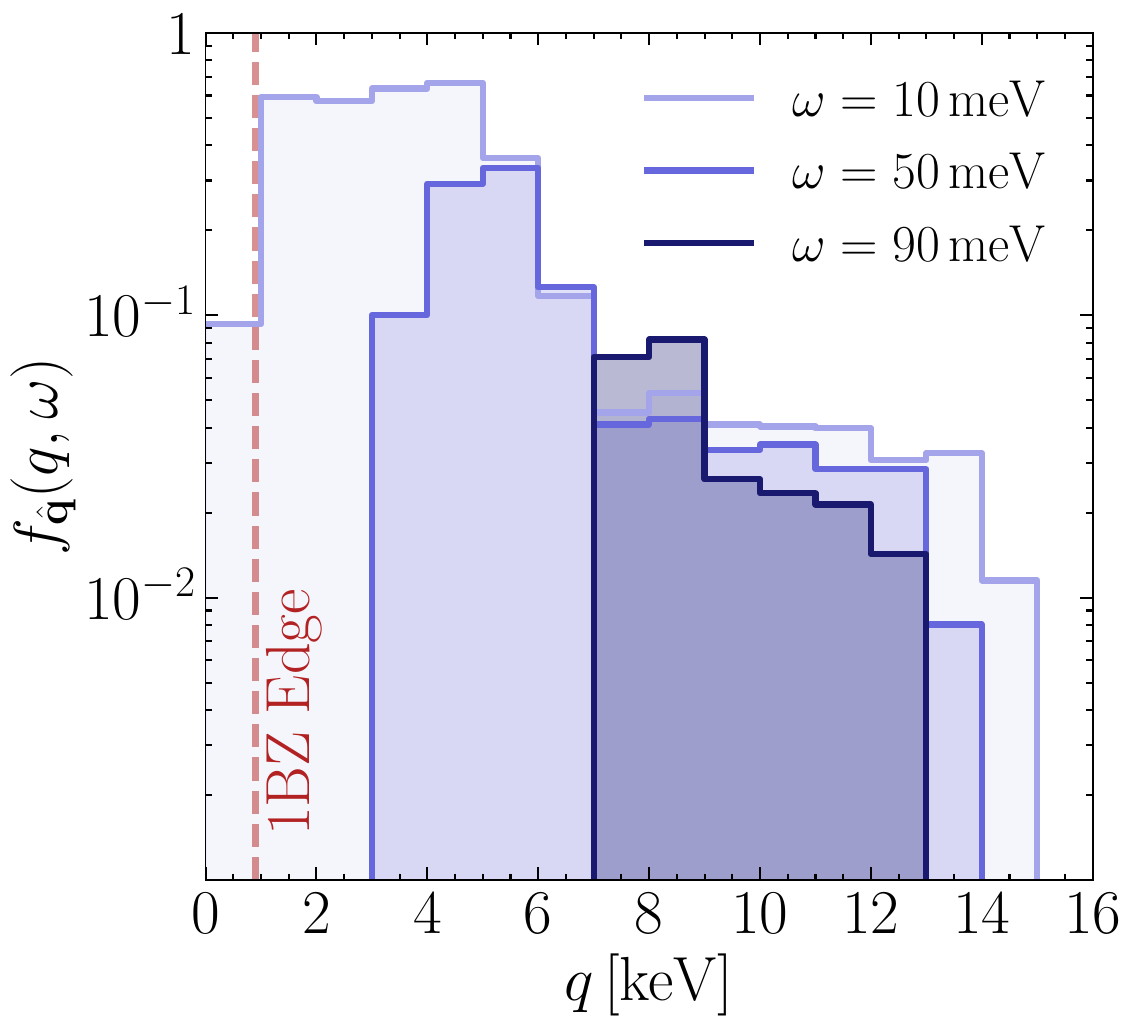}
    \caption{\textbf{Left:} The kinematic regions accessible in a light DM scattering event (gray), covered by the data of Ref.~\cite{princep2017full} (blue), and accessible by extrapolating the data downward in $q$ to the edge of the first Brillouin zone (red). \textbf{Right:} The fraction $f_{\hat{\mathbf{q}}}(q, \omega)$ of angles in $\qv$ in the data, as a function of $q$, for three values of $\omega$. For large $q$, only a small fraction of angles are covered.}
    \label{fig:data_kinematics_summary}
\end{figure*}

\subsubsection*{Dynamic Structure Factor}

Neutron scattering experiments typically report their results in terms of the ``dynamic structure factor'' $S_n(\qv, \omega)$. Though conventions vary, here we define 
\begin{align}
    \frac{d \sigma_n(\vv)}{d \Omega \, d E'} = \frac{|\v{p}'|}{|\v{p}|} \, S_n(\qv, \omega_{\qv}) \, ,
    \label{eq:diff_NS_cross_section}
\end{align}
where $\Omega$ is the solid angle of the outgoing neutron, other symbols are defined as in Sec.~\ref{subsec:suscep_scatter}, and $\sigma_n$ is the neutron scattering cross section per formula unit. The formula unit is the smallest representation of the chemical formula (\ce{Y3Fe5O12}), in contrast to the unit cell which is the smallest repeating structure. In YIG, the unit cell contains $N_\text{f} = 8$ formula units. Note that $\sigma_n$ is averaged over neutron spin, since the experimental measurements that we utilize employed an unpolarized neutron beam. 

The total scattering rate is $\Gamma_n = N \, N_\text{f} \, \sigma_n |\vv| / V$, where $N$ is the number of unit cells in the target. Integrating the differential cross section in \Eq{diff_NS_cross_section}, we find
\begin{align}
    \Gamma_n(\vv) = \frac{N_\text{f}}{m_n^2 \Omega_c} \int  \dd^3\qv ~ S_n(\qv, \omega_{\qv}) \, ,
\end{align}
where $\Omega_c = V/N$ is the volume of the unit cell. Comparing with \Eq{neutron_scattering_rate_sus}, we obtain
\begin{align}
    S_n(\qv, \omega) = \frac{1}{N_\text{f}} \frac{\gamma_n^2 e^4}{8 \,  m_e^2} \, \frac{\Omega_c}{(2 \pi)^3} \, P_T^{ij} \x''_{ij}(\qv, \omega_{\qv}) \, .
    \label{eq:S_chi_relationship}
\end{align}
Therefore, neutron scattering data directly determines $P_T^{ij} \x''_{ij}(\qv, \omega_{\qv})$. Referring to Table~\ref{tab:models}, we see that the magnetic dipole and anapole DM models also have form factors proportional to $P_T^{ij}$, implying that neutron scattering can be directly used to infer the corresponding DM-scattering rates. For now, we will focus on these two models; in Sec.~\ref{sec:other_ints}, we will present a physical interpretation of this fact, and explore how we can infer rates for other DM models. 

\subsection{Kinematic Coverage of Neutron Scattering Data}
\label{subsec:connection_to_data}

The maximum incoming velocity of DM in the detector frame is $v_\text{max} = v_\text{E} + v_\text{esc}$. As a result, the energy $\omega_{\qv}$ that DM can deposit in a scattering event, determined by \Eq{energy_deposit}, is bounded by $\omega_{\qv} \leq q \, v_\text{max} - q^2 / (2 m_\x)$. In addition, demanding that $\omega_{\qv} \geq 0$ implies an upper bound on the deposited momentum, $q \leq 2 m_\x v_{\text{max}}$, where the equality corresponds to backwards scattering. The kinematic region defined by these constraints is shaded gray in the left panel of \Fig{data_kinematics_summary}. To extract the DM scattering rate purely from data, the dynamic structure factor $S_n(\qv, \omega)$ should be measured for all values of $ \qv$ and $\omega$ in this region. 

In principle, a neutron beam with a speed of at least $v_{\text{max}} \sim 2 \times 10^{-3}$, corresponding to a $\keV$-scale kinetic energy, would be able to cover the full kinematic region for DM of any sub-GeV mass. However, as we will discuss in \Sec{neutron_temps}, detectors with $\meV$-scale energy resolution require much cooler neutrons, with speed $\sim 10^{-5}$. For instance, in the dataset of Ref.~\cite{princep2017full}, neutrons of energy $120 \, \meV$ were used. Such neutrons carry as much momentum as MeV mass DM. Though they carry significantly less energy, it is sufficient to probe all magnon excitations, as the maximum magnon energy is about $110 \, \meV$.

The set of all $\{q, \w\}$ covered by the data is shaded blue in \Fig{data_kinematics_summary}. Though the general shape of the region is determined by kinematics, it is also limited by some experimental factors. First, as discussed in Ref.~\cite{princep2017full}, the energy resolution at zero energy transfer is $5 \, \meV$, so data at $\w \lesssim 5 \, \meV$ is dominated by elastic scattering. Since these processes would not be visible in a DM detection experiment, we remove this part of the dataset. Second, the detectors in the MAPS spectrometer extend a maximum of $60^\circ$ out from the beam direction~\cite{MAPS}, so that backward recoils are not visible; this determines the upper-right boundary of the region. Finally, to avoid the beam, which has a typical angular spread of $\sim 1^\circ$, detectors are placed at least a few degrees away from the forward direction; this determines the lower-left boundary of the region. 

Referring to \Eq{energy_deposit}, when $\omega$ and $q$ are fixed, the value of $\qv \cdot \v{v}$ is fixed by energy conservation. To sample additional directions $\hat{\qv}$ of the momentum transfer, one can rotate the crystal, which in the crystal's frame corresponds to changing the direction $\vv$ of the incoming beam. In the experiment, the crystal was rotated about only a single axis through an arc of $\sim 120^\circ$. Thus, not all $\hat{\qv}$ were probed, particularly at higher deflection angles $\theta > 20^\circ$ where the detector covers only a narrow strip. 

To quantify this effect, we bin the raw data in a cubical grid with momentum spacing $\Delta q = 0.1 \, \text{keV}$ for all axes. For a given $q$ and $\omega$, we define $f_{\hat{\qv}}(q, \w)$ to be the fraction of the shell defined by $|\qv| \in [q - \delta q / 2, q + \delta q / 2]$ that is covered by bins in the data. In the right panel of \Fig{data_kinematics_summary}, we show this quantity for a shell of width $\delta q = 1 \, \text{keV}$. The coverage becomes worse at higher deflection angles, corresponding to higher $q$ for a given $\omega$. For further discussion, see App.~\ref{app:data}, where we provide visualizations of how the covered region corresponds to the detector geometry and crystal rotation.

To address these deficiencies in the data, we will extrapolate it in two simple ways:

\begin{enumerate}
\item \textbf{Angular Extrapolation.} For our angular extrapolation we approximate $S_n(\qv, \omega) \approx \bar{S}_n(q, \omega)$ for all $q$ and $\omega$, where $\bar{S}_n$ is defined by averaging over all $\hat{\qv}$ covered by the data for a given $q$ and $\omega$. This is justified because most missing angular data is at high $q$, well outside the first Brillouin zone (1BZ), which washes out the anisotropic details of the lattice. To understand this, note that the momentum vector can be decomposed as $\qv = \mathbf{k} + \mathbf{G}$ where $\mathbf{k}$ lies within the 1BZ and $\mathbf{G}$ is a reciprocal lattice vector. A small change in the direction of $\qv$ leads to a large change in $\mathbf{k}$, which controls much of the angular dependence of $S_n$. Then even an incomplete sampling of $\qv$ effectively samples over all $\mathbf{k}$, and features that depend on $\mathbf{k}$ are washed out, as shown in App.~\ref{app:data}. However, for $q \gtrsim 10 \, \text{keV}$ there can be $\mathcal{O}(1)$ magnetic form factor anisotropies~\cite{Marocco:2025eqw} which would break our isotropic approximation. We leave the inclusion of these details for future work.

\item \textbf{Magnitude Extrapolation.} After performing angular extrapolation, we can additionally extrapolate for smaller values of $q$. Here, we simply apply a constant extrapolation, i.e., we take $\bar{S}_n(q, \omega) = \bar{S}_n(q_\text{min}(\omega), \omega)$ for $q < q_\text{min}(\omega)$, where $q_\text{min}(\omega)$ is the minimum momentum transfer available in the data.\footnote{An alternative power-law extrapolation, i.e., $\bar{S}_n(q, \omega) = (q/q_\text{min})^\alpha \bar{S}_n(q_\text{min}, \omega)$ for $q < q_\text{min}$ with $\alpha$ fit to the available data, yielded similar sensitivity results.} This is a conservative assumption, as we generally expect $\bar{S}_n(q, \omega)$ to decrease at higher $q$ due to falling magnetic form factors~\cite{Marocco:2025eqw}, so that its value at lower $q$ should actually be higher than that used in our extrapolation. In addition, since we expect the dynamic structure factor to have non-trivial features for $q$ within the 1BZ, we conservatively stop the extrapolation at the edge of the 1BZ, $q_{1\text{BZ}} = 987 \, \text{eV}$, as indicated in the left panel of \Fig{data_kinematics_summary}.
\end{enumerate}

These extrapolation schemes are simple and conservative, but in future work one could use more refined methods to get more accurate results. For instance, to improve the angular extrapolation, one could fit an angular dependence compatible with the symmetries of the crystal lattice, and explicitly accounting for the magnetic form factors would improve both the angular and magnitude extrapolations. Alternatively, data with more complete angular and magnitude coverage could be taken to remove the need for extrapolation entirely.

\subsection{Sensitivity Projections}
\label{subsec:sensitivity_proj}

We now compute the signal-limited sensitivity to $g_{\x} g_e$ for the magnetic dipole and anapole DM models, where $g_\x$ and $g_e$ are the dimensionless DM and electron coupling constants, respectively, as defined in Table~\ref{tab:models}. For concreteness, we will always assume a kg-yr exposure and an experimental energy threshold of $\omega_\text{th} = 10 \, \text{meV}$. We assume no backgrounds, so that a 95\% C.L.~sensitivity corresponds to $3$ expected scattering events.

\subsubsection*{Numeric Computation of the Scattering Rate}

We begin by explicitly showing how our extrapolation schemes are numerically implemented, and their effect on the projected sensitivity. For concreteness, we will focus on magnetic dipole DM coupled through a light dark photon; as shown in Table~\ref{tab:models}, the anapole DM case is simply related by rescaling the form factor by $|\qv|^2 / (4 m_\x^2)$. 

For magnetic dipole DM, an analogous calculation to that of Sec.~\ref{subsec:ns_derivation} gives a scattering rate of
\begin{align}
    \Gamma(\vv) & = \frac{g_\x^2 g_e^2}{m_\x^2 \gamma_n^2 e^4} \frac{N_\text{f}}{\Omega_c} \int \dd^3\qv ~ S_n(\qv, \omega_{\qv}) \, ,
\end{align}
where we have written this in terms of the dynamic structure factor using \Eq{S_chi_relationship}. The scattering rate per unit target mass, averaged over initial DM velocity, is therefore
\begin{align}
    R = \frac{\rho_\x}{2 \pi \rhoT} \frac{g_\x^2 g_e^2}{m_\x^3 \gamma_n^2 e^4}  \frac{N_\text{f}}{\Omega_c} \int \dd \omega \, \dd^3\qv ~ g(\qv, \omega)\, S_n(\qv, \omega)  \, ,
    \label{eq:R_MDM}
\end{align}
where the function $g(\qv, \omega)$ is given by~\cite{Trickle:2019nya}
\begin{align}
    g(\qv, \omega) & \equiv 2 \pi \int \dd^3 \vv ~ f_{\x}(\vv) \, \delta(\w - \w_\qv) \, .
\end{align}

\begin{figure}[t!]
    \centering
    \includegraphics[width=\linewidth]{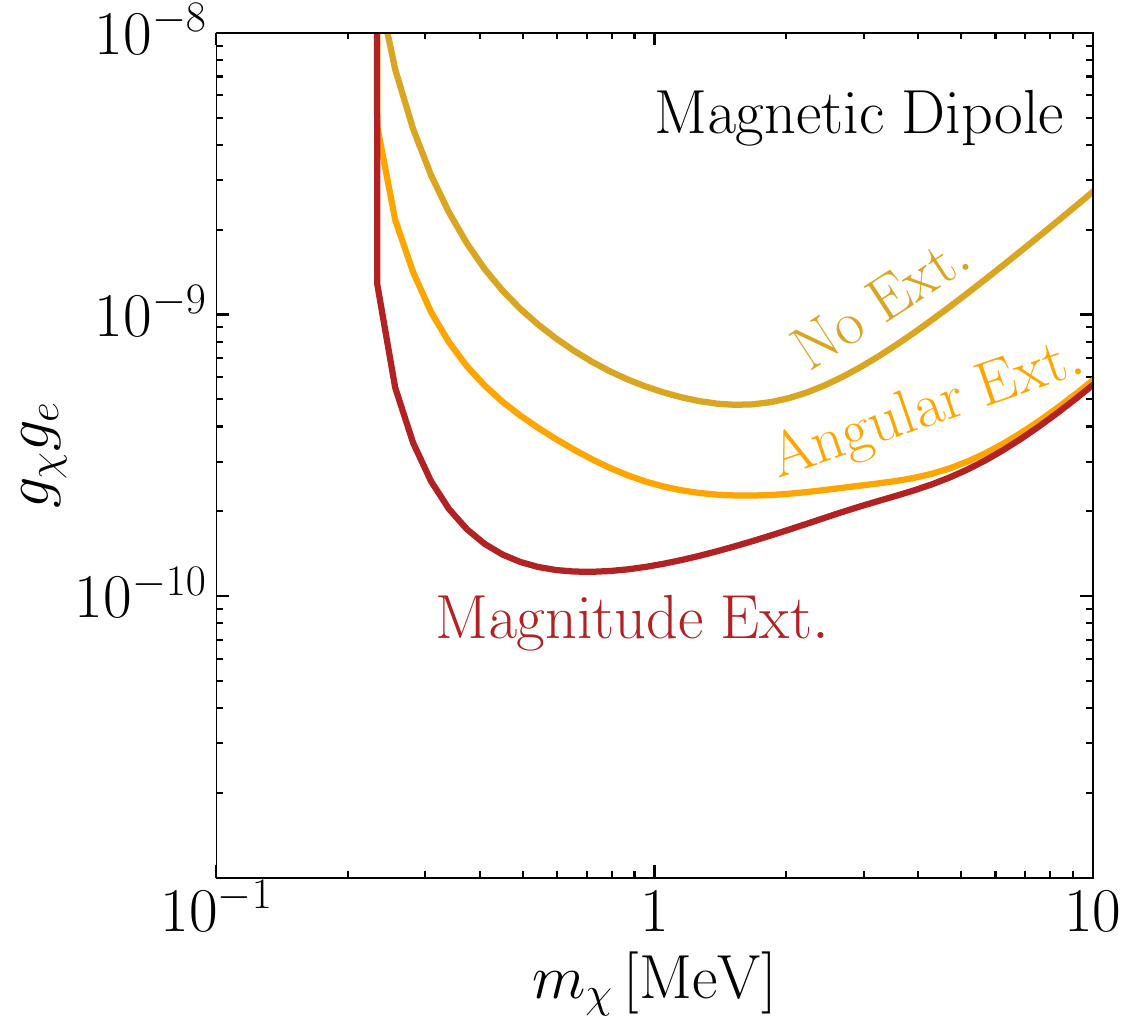}
    \caption{The calculated sensitivity of YIG to magnetic dipole DM, for various extrapolation schemes used in our data-driven approach. With ``no extrapolation,'' only the existing data is used. With ``angular extrapolation,'' we correct for missing angular coverage (shown in the right-panel of \Fig{data_kinematics_summary}) by constructing an angular averaged dynamic structure factor $\bar{S}_n(q, \w)$. With ``magnitude extrapolation,'' we fill in the red region, as shown in the left-panel of \Fig{data_kinematics_summary}, by assuming $\bar{S}_n(q, \w)$ remains constant as $q$ decreases. Other assumptions are as in \Fig{sensitivity_compare}.}
    \label{fig:reach_extrapolations}
\end{figure}

To determine $S_n (\qv , \w)$, we bin the neutron scattering data using the publicly available code \textsc{horace}~\cite{ewings2016horace}, resulting in a grid with momentum spacing $\Delta q = 0.1 \, \text{keV}$ and energy spacing $\Delta \omega = 1\, \text{meV}$.\footnote{The raw data was processed in Mantid~\cite{arnold2014mantid} and provided by Ref.~\cite{russellpriv} along with access to the ISIS Data Analysis Service~\cite{ISIS}.} Then the integral in \Eq{R_MDM} can be discretized to 
\begin{align}
    R \approx &\frac{\rho_\x}{2 \pi \rhoT} \frac{g_\x^2 g_e^2}{m_\x^3 \gamma_n^2 e^4} \frac{N_\text{f} \, (\Delta q)^3 \Delta \omega}{\Omega_c}  
    \nl
    & \times \sum_{\v{q}, \omega} g(\qv, \omega)\, S_n(\v{q}, \omega) \, ,
    \label{eq:direct}
\end{align}
where the sum runs over the four-dimensional grid. This gives the unextrapolated result. 

To implement angular extrapolation, we define an angular-averaged dynamic structure factor $\bar{S}_n(q, \omega)$, as motivated in Sec.~\ref{subsec:connection_to_data}. In terms of this isotropic structure factor, the angular integral in \Eq{R_MDM} can be evaluated to yield 
\begin{align}
    R \approx \frac{2 \pi \rho_\x}{\rhoT} \frac{g_\x^2 g_e^2}{m_\x^3 \gamma_n^2 e^4} \frac{N_\text{f}}{\Omega_c} \int \dd q \, \dd\omega ~ q\, \eta(q, \omega)\, \bar{S}_n(q, \omega) \, ,
    \label{eq:R_MDM_uniform}
\end{align}
where $\eta(q, \omega)$ is given by~\cite{Trickle:2019nya}
\begin{align}
    \eta(q, \omega) \equiv \int \dd^3\v{v} ~ \frac{f_\x(\v{v})}{v} \, \Theta(v - v_\text{min})
\end{align}
and $v_\text{min} = |\qv| / (2 m_\x) + \omega / |\qv|$. To compute $\bar{S}_n(q, \omega)$, we discretize it into bins of width $\delta q = 0.3 \, \text{keV}$ in $q$. The value of $\bar{S}_n(q, \omega)$ for each bin in $q$ is calculated by averaging over all $S_n(\v{q}, \omega)$ in the data for which $q - \delta q / 2 \leq |\v{q}| \leq q + \delta q / 2$. Then the integral in \Eq{R_MDM_uniform} is discretized to
\begin{align}
    R \approx \frac{2 \pi \rho_\x}{\rhoT} \frac{g_\x^2 g_e^2}{m_\x^3 \gamma_n^2 e^4} \frac{N_\text{f}}{\Omega_c} \delta q \, \Delta \omega\sum_{q,\omega} q \, \eta(q, \omega)\, \bar{S}_n(q, \omega) \, ,
    \label{eq:R_MDM_uniform_discrete}
\end{align}
where the sum runs over all the $q$ and $\omega$ in the angularly averaged dataset. To additionally implement magnitude extrapolation, we simply extend the sum down in $q$, setting $\bar{S}_n(q, \omega) = \bar{S}_n(q_\text{min}(\omega), \omega)$ for $q_{\text{1BZ}} < q < q_\text{min}(\omega)$.

The resulting sensitivities are shown in \Fig{reach_extrapolations}. Angular extrapolation yields a significantly large signal rate, particularly at high $m_\x$, as expected from the falloff in angular coverage as shown in the right-panel of \Fig{data_kinematics_summary}. Similarly, magnitude extrapolation further enhances the signal rate, particularly at lower $m_\x$. This is because the form factor for magnetic dipole DM rapidly grows with $q$, so that the scattering rate is dominated by the highest $q$ allowed by kinematics; for lighter DM, much of this region is only covered by extrapolation, as shown in the left-panel of \Fig{data_kinematics_summary}. In all cases, the sensitivity rapidly falls for $m_\x \lesssim 200 \, \keV$. For these masses, DM scattering takes place entirely within the 1BZ, where there is no extrapolation and little data.

\subsubsection*{Comparison to Previous Results}

\begin{figure*}[ht!]
    \centering
    \includegraphics[width=\linewidth]{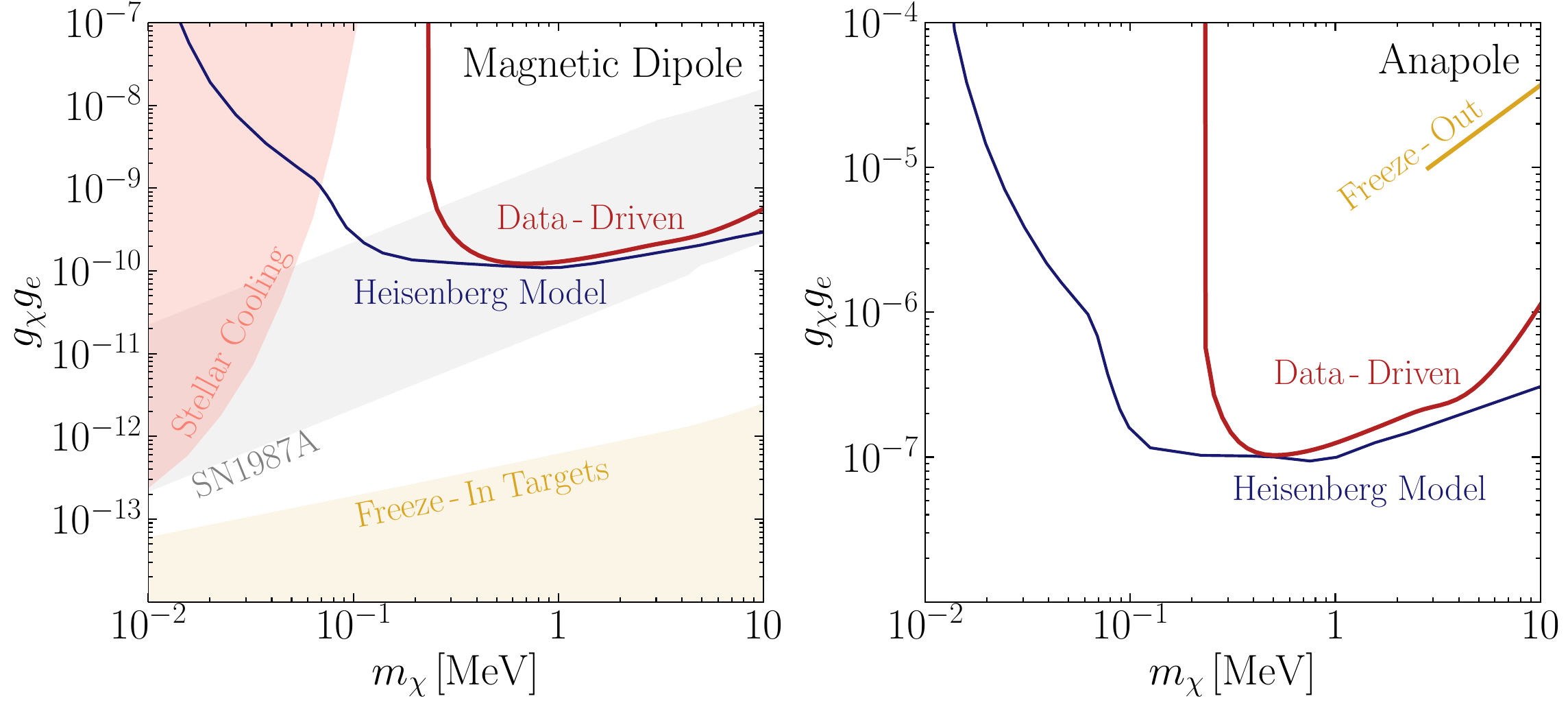}
    \caption{Projected sensitivity of a YIG target to magnetic dipole (left) and anapole (right) DM with a light mediator (see Table~\ref{tab:models}). We compute the 95\% C.L.~sensitivity assuming a kg-year exposure, no background events, and an experimental energy threshold of $\omega_\text{th} = 10 \, \text{meV}$. The ``data-driven" curves are the main result of this work. They are computed using the raw neutron scattering data of Ref.~\cite{princep2017full}, using both angular and magnitude extrapolations (see \Fig{reach_extrapolations}). The ``Heisenberg model'' curves, taken from Ref.~\cite{Trickle:2019ovy}, were computed by fitting a theoretical model to the same dataset. We also show astrophysical constraints and freeze-out and freeze-in targets, collected from Refs.~\cite{Chang:2019xva,Chu:2019rok}. The freeze-in region in the left-panel assumes the DM is produced by the magnetic dipole interaction, and that the reheat temperature was at least $10 \, \text{MeV}$.}
    \label{fig:sensitivity_compare}
\end{figure*}

In \Fig{sensitivity_compare}, we compare the sensitivity computed with the data-driven approach discussed here to the first principles calculation pioneered in Ref.~\cite{Trickle:2019ovy}, for both magnetic dipole and anapole DM. Note that the first principles calculation also relied on the neutron scattering data of Ref.~\cite{princep2017full}, albeit indirectly. In that work, the target was modeled with the Heisenberg Hamiltonian, the data was used to compute best fit parameters, and magnon modes were found by diagonalizing the Hamiltonian, as in Ref.~\cite{Toth_2015}.

The results are in close agreement for DM masses near $m_\x \sim 1 \, \text{MeV}$. For smaller $m_\x$, the data-driven sensitivity falls rapidly because the data does not cover the relevant DM scattering kinematic region. For larger $m_\x$, both results have limitations. The Heisenberg model neglects the magnetic form factor of $\text{Fe}^{3+}$, which suppresses the scattering rate by an order of magnitude for high momentum transfers, $q \gtrsim 10 \, \text{keV}$~\cite{boothroyd2020,matthewman1982cambridge}. The data-driven method automatically captures this suppression. However, as discussed in App.~\ref{app:subdominant_ints}, at high $q$ it also includes an increasing number of scattering events that produce phonons, and thus overestimates the magnon scattering rate. We leave a detailed treatment of the high $m_\x$ regime, which requires subtracting the phonon contribution, for future work.

In \Fig{sensitivity_compare}, we also show other astrophysical bounds and DM relic abundance targets, collected from Refs.~\cite{Chang:2019xva,Chu:2019rok}. However, we note that these bounds have been recast from those assuming the light mediator is the photon. This is appropriate for processes dominated by diagrams involving direct $\x \x \leftrightarrow ee$ interactions (such as DM production via $s$-channel electron annihilations), but not appropriate for bounds which rely on an external photon or plasmon, such as LEP missing energy searches using photons~\cite{Chu:2019rok}. The relic abundance targets and shaded regions shown in \Fig{sensitivity_compare} are all, dominantly, due to the former set of processes. Finally, we note that these bounds depend on the specific ultraviolet completion of the DM model. For instance, if the energy in a process exceeds the scale beyond which the Lagrangians in Table~\ref{tab:models} are no longer valid, then any derived bounds may become inapplicable. This is most relevant for bounds derived from Big Bang nucleosynthesis (BBN), as they involve temperatures $\sim (1 - 10) \, \MeV$ greater than most of the DM masses considered. Furthermore, in certain models~\cite{DeRocco:2020xdt,Masso:2006gc}, the in-medium DM-SM coupling is screened in higher density and temperature environments. Hence, we do not explicitly consider BBN limits here.

\section{Discussion}
\label{sec:discussion}

In the previous section, we focused on one particular dataset, and its application to magnetic dipole and anapole DM. Here we consider the additional information one could gain from other types of scattering experiments, and how the rate could be inferred for other DM models. 

\subsection{Other Neutron Scattering Experiments}
\label{sec:neutron_temps}

The dataset considered in this work was collected with neutrons of incident energy $E_i \approx 120 \, \meV$, because this was sufficient to cover the full magnon spectrum of YIG with good energy resolution. We have seen in Sec.~\ref{subsec:connection_to_data} that this is a decent, but not perfect kinematic match for DM scattering. These issues can be partially addressed by varying the neutron energy. Colder neutrons could probe the lower momentum transfers relevant for sub-$\MeV$ DM, while hotter neutrons could probe the higher energy transfers relevant for $\MeV$ DM. 

For the MAPS time-of-flight neutron spectrometer, the full range of available neutron energies is $15 \, \meV \leq E_i \leq 2 \, \eV$~\cite{MAPS}. By using higher energy neutrons, we would be able to sample higher $q$ and $\omega$, albeit at the cost of a proportionally worse energy resolution~\cite{MAPS,boothroyd2020}. On the other hand, experiments with cold neutrons can achieve exceptional energy resolution. For example, neutron spin-echo spectroscopy can be used to probe energy transfers with $\sim \mu\eV$ resolution~\cite{mezei2002neutron,keller2022neutron}. 

In addition, to derive an accurate result at higher DM masses, $m_\x \gtrsim \text{few} \, \text{MeV}$, the phonon background must be subtracted. This can be done by theoretical modeling, or more directly by spin-polarized neutron scattering measurements, as magnetic scattering preferentially flips the neutron's spin. 

\subsection{Photon and Electron Scattering}

A key advantage of expressing the spin-dependent DM scattering rate in terms of the spin susceptibility is that the latter can be measured using any probe that couples to electron spins. Therefore, in principle one could also infer it using magnetic X-ray or electron scattering. However, while magnetic scattering is the dominant electromagnetic interaction for a neutron, it is generically subdominant for X-rays and electrons. For X-rays of energy $E$, the ratio of the cross sections for magnetic and charge scattering is of order $(E/m_e)^2$~\cite{lovesey1987magnetic}, which is $\sim 10^{-4}$ for a $E = 5 \, \keV$. For electrons of typical speed $v$, magnetic interactions are suppressed relative to the Coulomb interaction by $\order{v^2}$ in the nonrelativistic limit. 

Thus, for both cases one must focus on special regimes where the magnetic interaction dominates. In resonant inelastic X-ray scattering (RIXS)~\cite{ament2011resonant}, the X-ray energy is tuned to an absorption edge. This enhances magnetic scattering, and allows magnon spectra to be measured~\cite{haverkort2010theory}. In spin-polarized electron energy loss spectroscopy~\cite{vollmer2003spin}, the spins of the electrons are measured, since magnetic scattering dominantly produces spin flip events. 

A key obstacle to applying these methods to DM scattering is achieving sufficient energy resolution. Compared to neutrons, electrons have a much larger kinetic energy for a given momentum, so energy measurements must have a much higher relative precision to reach the same absolute precision. Typical instruments for electron energy loss spectroscopy have an energy resolution of order $100 \, \meV$~\cite{ROTH201485}, and it was noted in Ref.~\cite{Boyd:2022tcn} that this presents an obstacle to using it to measure even the dielectric function, in the regime relevant for light DM scattering. However, newer instruments may reach an energy resolution of about $5 \, \meV$, competitive with neutron scattering~\cite{krivanek2019progress}. 

Similarly, to resolve a $\meV$ energy transfer in a $\keV$ X-ray, one needs a very high resolving power of $10^6$. As such, RIXS has historically had much worse energy resolution than neutron scattering, but new instruments have reached resolutions of tens of $\meV$~\cite{moretti2018high,brookes2018beamline,zhou2022i21}. If magnetic X-ray scattering can be measured with sufficient resolution, then it yields complementary information to neutron scattering. X-rays can probe higher energy transfers, and resolve lower momentum transfers. 

A potential unique advantage of magnetic X-ray scattering is that it couples differently to spin and orbital angular momentum, allowing their contributions to the magnetization to be decomposed, while neutron scattering only couples to the total magnetization. This ability might be relevant for DM models which couple selectively to electron spin. However, it was not relevant here, as the $3d$ electrons in transition metal compounds such as YIG experience strong crystal fields which quench their orbital angular momentum~\cite{Trickle:2019ovy,Mitridate:2020kly,boothroyd2020}. 

\subsection{Other Dark Matter Interactions}
\label{sec:other_ints}

We found that for magnetic dipole and anapole DM, the scattering rate only depends on the combination 
\be
P_T^{ij} \x_{ij}'' = \text{tr} \, \x'' - \hat{\qv} \cdot \x'' \cdot \hat{\qv}
~,
\ee
which contains a trace over the components of $\x''$ transverse to $\qv$. The presence of a trace reflects our assumption that the DM was unpolarized. If DM were spin polarized, the rate could still be determined using spin-polarized neutron scattering data. However, for these models, the rate would still only depend on the $4$ components of $\x_{ij}''$ transverse to $\qv$, rather than all $9$.

This fact has a simple physical interpretation. Referring to \Eq{linear_response}, the field $\v{\Phi}$ stands for a spin-coupled effective magnetic field. The magnetic field of a neutron has no divergence; likewise, for magnetic dipole and anapole DM we have $\qv \cdot \bm{\mathcal{O}}(\mathbf{q}, \omega) = 0$. Furthermore, in all of these cases the scattered particle couples to the current $\v{J}$ induced in the medium. Since $\v{J} \propto \nabla \times \v{s}_e(\mathbf{x}, t)$, it does not depend on the divergence $\qv \cdot \v{s}_e(\mathbf{q}, \omega)$. This is a well-known limitation of neutron scattering experiments~\cite{boothroyd2020}, which implies that they cannot directly measure the full magnetization profile. Therefore, for these models, only $4$ of the components of the susceptibility defined in \Eq{linear_response} contribute to scattering.

By contrast, the DM models in the bottom three rows of Table~\ref{tab:models} yield rates proportional to the longitudinal component $\hat{\qv} \cdot \x'' \cdot \hat{\qv} \equiv \x''_{qq}$. This has a similar physical interpretation. For the scalar mediator models, the effective magnetic field acting on the electron spins is $\v{\Phi} \propto \bm{\nabla} \phi$, which has a divergence but no curl. For the dark electron EDM model, $\v{\Phi} \propto \v{E}'$ where the dark electric field $\v{E}'$ of a nonrelativistic DM particle has divergence but no curl. 

In general, $\x''_{qq}$ is an independent material quantity which cannot be inferred from $P_T^{ij} \x_{ij}''$ alone. However, in many cases they can be related. For example, a magnetic material comprised of many domains oriented in different directions would be isotropic, so that $\x''_{ij} \propto \delta_{ij}$, implying 
\begin{equation}
\x''_{qq} = \frac12 P_T^{ij} \x''_{ij}.
\end{equation}
This possibility was previously pointed out in Ref.~\cite{Marocco:2025eqw}.

Alternatively, most magnets are collinear, i.e., each spin is classically parallel or antiparallel to a single direction.\footnote{Noncollinear magnetic order can be produced by, e.g., strong spin-orbit couplings or frustrated exchange interactions.} For a collinear magnet with a single domain magnetized along $\hat{\v{z}}$, rotational symmetry about the $z$-axis implies that $\x''_{xx} = \x''_{yy}$ and $\x''_{xy} = -\x''_{yx}$~\cite{lovesey1984theory}. Moreover, as discussed further in App.~\ref{app:spin_resp_model}, all other components vanish. These constraints imply that
\begin{align}
\x''_{qq} &= (1 - \hat{\qv}_z^2) \, \x''_{xx}, \label{eq:collinear_1} \\
P^{ij}_T \x_{ij}'' &= (1 + \hat{\qv}_z^2) \, \x''_{xx} \label{eq:collinear_2}
\end{align}
which allows $\x''_{qq}$ to be inferred from $P^{ij}_T \x_{ij}''$. These results apply to the dataset used in this work, because YIG is a collinear magnet and the sample was magnetized to saturation. Thus, in principle we may infer the DM scattering rate in YIG for any of the models in Table~\ref{tab:models}.

As long as a magnet is not strongly anisotropic, the angular dependence of elements of its susceptibility, such as $\x_{xx}''$, should be largely washed out at large momentum transfers. However, most of the DM models we consider give scattering rates proportional to either $\x''_{qq}$ or $P^{ij}_T \x_{ij}''$, both of which explicitly depend on $\hat{\qv}_z$. The sole exception is the axial vector mediator model, which depends on the $\text{tr} \, \x'' = \x''_{qq} + P^{ij}_T \x_{ij}'' = 2 \, \x''_{xx}$. Since the typical $\hat{\qv}_z$ depends on the relative orientation of the DM momentum transfer and the magnet's axis of symmetry, these models will display daily modulations of the scattering rate, which can help distinguish signal from background~\cite{Trickle:2019nya,Coskuner:2019odd,Coskuner:2021qxo,Boyd:2022tcn,Marocco:2025eqw,Stratman:2024sng}, and distinguish between DM models.

\section{Conclusion}

The direct detection of light DM depends in detail on how novel target materials and excitations respond to DM interactions. In this work we have shown how electron spin-dependent interaction rates can be connected to the magnetic susceptibility, which can be measured using neutron scattering. We have demonstrated the power of this approach with currently available data, and discussed how it could be further improved with future measurements. 

We have focused on a YIG target for concreteness. However, a key advantage of our data-driven approach is that it allows one to find the spin-dependent DM interaction rate in a wide range of target materials using a single experimental method. For instance, ferrites~\cite{Berlin:2023ubt}, anti-ferromagnets~\cite{Esposito:2022bnu,Catinari:2024ekq}, and quasi-two-dimensional magnets~\cite{Marocco:2025eqw} have all been considered in the DM literature. In addition, rare earth magnets may be useful at higher DM masses, since the magnetic form factor of an $f$ ion falls more slowly at high momentum transfer. Neutron scattering can pinpoint the materials with the highest overall sensitivity and the strongest daily modulation in the signal rate, and thereby accelerate the discovery of dark matter.

\acknowledgments

We thank Zachary Bogorad, Bill Gannon, Giacomo Marocco, Shashin Pavaskar, and Ryan Plestid for discussions. We especially thank Russell Ewings for access to the neutron scattering data. KZ was supported by the Office of High Energy Physics of the U.S. Department of Energy under contract DE-AC02-05CH11231. TT was supported by Quantum Information Science Enabled Discovery 2.0 (QuantISED 2.0) for High Energy Physics (DE-FOA-0003354). This manuscript has been authored by Fermi Forward Discovery Group, LLC under Contract No. 89243024CSC000002 with the U.S. Department of Energy, Office of Science, Office of High Energy Physics. 

\bibliographystyle{utphys3}
\bibliography{MagELF}

\clearpage
\appendix
\onecolumngrid

\section{Scattering Rate Calculations}
\label{app:calcs}

\subsection{Derivation of the Scattering Potential}
\label{app:calcs_potential}

The interaction Hamiltonian density $\mathcal{H}_e(\mathbf{x}, t)$ defined in \Eq{H_int_density} involves only the electron field, and therefore only acts on the electron degrees of freedom; all the DM information is encoded in the external potential $\Phi(\mathbf{x}, t)$. We construct this effective interaction by matching the effective interaction Hamiltonian, $H_{e} = \int \, \dd^3 \mathbf{x} \, \mathcal{H}_{e}(\mathbf{x}, t)$ to the matrix element of the entire DM-electron scattering process, without including the electronic states. That is, we set
\begin{align}
    -i \int \dd t \, H_{e} \equiv \frac{\langle \mathbf{p}', s' | \,  i \mathcal{T} \, | \mathbf{p}, s \rangle}{ \sqrt{ \langle\mathbf{p}', s' | \mathbf{p}', s' \rangle \langle\mathbf{p}, s | \mathbf{p}, s \rangle }}
    \label{eq:H_int_def}
\end{align}
where the DM states are as defined in Sec.~\ref{subsec:suscep_scatter}. As usual~\cite{Peskin:1995ev}, the transfer matrix $\mathcal{T}$ is defined by $1 + i\mathcal{T} \equiv T\{ e^{i \int \dd t \,\dd^3\mathbf{x} \, \mathcal{L}} \}$, where $\mathcal{L}$ is the electron-DM interaction Lagrangian and $T$ is the time-ordered product. The state normalization factors ensure $H_{e}$ has mass dimension one, independent of normalization convention. The definition in \Eq{H_int_def} ensures that the matrix element of $-i \int \dd t \, H_{e}$ between the initial and final electronic states matches the results of the full matrix element calculation.

For illustration, let us work through a simple concrete example. Consider the nonrelativistic interaction Lagrangian
\begin{align}
    \mathcal{L} \supset V^i \left( g_\x  s_\x^i + g_e \, s_e^i \right) \, ,
    \label{eq:L_int_example}
\end{align}
where $s_\x^i \equiv \psi_\x^\dagger \sigma^i \psi_\x$ is twice the $\x$ spin density, $V^i$ is a vector mediator with mass $m_V$, and $g_\x$ and $g_e$ are couplings. Then the right hand side of \Eq{H_int_def} is
\begin{align}
    \frac{\langle \mathbf{p}', s' | i \mathcal{T} | \mathbf{p}, s \rangle}{\sqrt{ \langle\mathbf{p}', s' | \mathbf{p}', s' \rangle \langle\mathbf{p}, s | \mathbf{p}, s \rangle}} & \approx -\frac{g_\x g_e}{2 m_\x V} \int \dd^4 y \, \dd^4 x \, \langle \mathbf{p}', s' | \, T\left\{ s_\x^i(y) \,  V^i(y)  V^j(x) \, s_e^j(x) \right\} | \mathbf{p}, s \rangle  \nonumber \\ 
    & \approx -\frac{g_\x g_e}{2 m_\x V} \int \dd^4 y \, \dd^4x \, \left[ 2m_\x \sigma^i_{ss'} e^{i (p' - p) \cdot y} \right]  D_V^{ij}(y - x) \, s_e^j(x) \nonumber \\
    & \approx i\frac{g_\x g_e}{V} \frac{1}{|\qv|^2 + m_V^2} \int \dd t \, \dd^3 \mathbf{x} \, \sigma_{ss'}^i e^{- i q \cdot x} \, s_e^i(\mathbf{x}, t) \, ,
    \label{eq:app_T_deriv}
\end{align}
to leading order in $g_\x$ and $g_e$, where $x^\mu = (t, \mathbf{x})$, $y^\mu = (t', \mathbf{y})$, and $q^\mu \equiv p^\mu - {p'}^\mu$. Above, the DM states are relativistically normalized, $\langle \mathbf{p}, s | \mathbf{p}, s \rangle = 2 m_\x V$ where $V$ is the target volume, $D_V^{ij}(q)~\approx~-i \delta^{ij} / (|\qv|^2 + m_V^2)$ is the Fourier transform of the $V^i$ propagator in the scattering kinematic limit $\omega_\qv \ll |\mathbf{q}|$, and $\sigma^i_{ss'} \equiv \xi_{s'}^\dagger \sigma^i \xi_s$, where $\xi_\uparrow = (1~~0)^T$ and $\xi_\downarrow = (0~~1)^T$. From \Eq{app_T_deriv} we can read off 
\begin{align}
\mathcal{H}_e = -\frac{g_\x g_e}{V} \frac{e^{-i q \cdot x}}{|\qv|^2 + m_V^2} \, \bm{\sigma}_{ss'} \cdot \mathbf{s}_e(\mathbf{x}, t),
\end{align}
which immediately yields the scattering potential $\bm{\Phi}(\mathbf{x}, t)$, which in turn corresponds to
\begin{align}
\mathcal{O}^i = \frac{g_\x g_e}{|\qv|^2 + m_V^2} \, \sigma_{ss'}^i \, . \label{eq:Phi_example_scatter}
\end{align}
Finding the external potential for any other process, including absorption processes, follows the same method shown here: match the (normalized) transition matrix element evaluated between the DM states to an interaction Hamiltonian.

\subsection{Derivation of the Scattering Rate}
\label{app:rates}

Here, we derive the scattering rate given in \Eq{scattering_rate}. Defining the interaction Hamiltonian by $H_{e}(t) = \int \dd^3 \mathbf{x} \, \mathcal{H}_{e}(\mathbf{x}, t)$, the leading order transition rate for the process $| I \rangle  = | \mathbf{p}, s\rangle \otimes | e \rangle \rightarrow | F \rangle = | \mathbf{p}', s' \rangle \otimes | e' \rangle$ is
\begin{align}
    \Gamma_{I \rightarrow F} & = \lim_{T \rightarrow \infty} \frac{1}{T} \left| \int_{-T/2}^{T/2} \dd t \,  \langle e' | H_{e} | e \rangle \right|^2 \, ,
    \label{eq:app_rate_1}
\end{align}
where $T$ is the observation time. Note that $H_{e}$ contains a matrix element involving the initial and final DM states, as shown in \Eq{H_int_def}, so that it can depend on $\mathbf{p}$, $\mathbf{p}'$, $s$, and $s'$. Expanding out $H_{e}$ using the first line of \Eq{dm_op_def}, we have
\begin{align}
    \Gamma_{I \rightarrow F} & = \lim_{T \rightarrow \infty} \frac{1}{V^2 T} \left| \mathcal{O}^i \int_{-T/2}^{T/2} \dd t \, e^{-i \omega_{\mathbf{q}} t} \int \dd^3 \mathbf{x} \, e^{i \mathbf{q} \cdot \mathbf{x}} \, \langle e' | s_{e}^i(\mathbf{x}, t) | e \rangle \right|^2 \nonumber \\ 
    & = \lim_{T \rightarrow \infty} \frac{1}{V^2 T} \left| \mathcal{O}^i \int_{-T/2}^{T/2} \dd t \, e^{-i \left( \omega_{\mathbf{q}} - (E_{e'} - E_e) \right) t} \int \dd^3 \mathbf{x} \, e^{i \mathbf{q} \cdot \mathbf{x}} \, \langle e' | s_{e}^i(\mathbf{x}, 0) | e \rangle \right|^2 \nonumber \\
    & = \frac{1}{V^2} \, \mathcal{O}_i \mathcal{O}_j^* \,  \langle e' | \tilde{s}_{e}^i(-\mathbf{q}) | e \rangle \, \langle e | \tilde{s}_{e}^j(\mathbf{q}) | e' \rangle \, 2 \pi \, \delta\left(\omega_{\mathbf{q}} - (E_{e'} - E_e) \right) \, ,
    \label{eq:app_rate_2}
\end{align}
where $\omega_{\mathbf{q}}$ is given by \Eq{energy_deposit}, and $\tilde{s}_e^i(\mathbf{q}) \equiv \int \dd^3 \mathbf{x} \, e^{-i \mathbf{q} \cdot \mathbf{x}} \, s_e^i(\mathbf{x}, 0)$. The total spin-averaged scattering rate per incoming DM particle is found by summing \Eq{app_rate_2} over all possible transitions, averaging over initial spins. This yields
\begin{align}
    \Gamma & = \frac{2 \pi}{2 S_\x + 1} \sum_{ss'} \sum_{\mathbf{p}'} \sum_{ee'} \frac{1}{V^2} \mathcal{O}_i \mathcal{O}_j^* \,  \langle e' | \tilde{s}_{e}^i(-\mathbf{q}) | e \rangle \, \langle e | \tilde{s}_{e}^j(\mathbf{q}) | e' \rangle \,  \delta\left(\omega_{\mathbf{q}} - (E_{e'} - E_e) \right) \nonumber \\
    & = \frac{\pi}{V} \int \frac{\dd^3\mathbf{q}}{(2 \pi)^3} \, \mathcal{F}^{ij}(\mathbf{q}, \omega_{\mathbf{q}}) \, \left[  \sum_{ee'} \langle e' | \tilde{s}_{e}^j(-\mathbf{q}) | e \rangle \, \langle e | \tilde{s}_{e}^i(\mathbf{q}) | e' \rangle  \,  \delta\left(\omega_{\mathbf{q}} - (E_{e'} - E_e) \right) \right] \, ,
    \label{eq:app_rate_3}
\end{align}
where the form factor $\mathcal{F}^{ij}(\mathbf{q}, \omega_{\mathbf{q}})$ is as defined in \Eq{F_def}. Applying \Eq{chi_pp} yields \Eq{scattering_rate} of the main text. The absorption rate formula \Eq{abs_rate} is derived similarly, though in that case the sum over final DM states is trivial.

\section{The Dynamical Spin Magnetic Susceptibility}

\subsection{Relation to Electronic States}
\label{app:suscep}

Here we derive \Eq{chi_pp} of the main text. The momentum-space susceptibility is given by the Kubo formula~\cite{kubo2012statistical}, 
\begin{equation} \label{eq:kubo_1}
\x_{ij}(\qv, \w) = \frac{i}{V} \int \dd^3 \v{x} \, \dd^3 \v{y} \, e^{- i \qv \cdot ( \v{x} - \v{y})} \int_0^\infty \dd t \, e^{i \w t} \sum_{e} p_e\, \langle e| \, [s_e^i(\v{x}, t), s_e^j(\v{y}, 0)] \, | e  \rangle \, ,
\end{equation}
where $p_e$ is the probability of occupancy of the initial state $|e \rangle$. Note that in the limit of a translationally invariant system, the integrand in \Eq{kubo_1} only depends on $\v{x} - \v{y}$, and the number of spatial integrals can be reduced to one. Using the fact that $s_e^i(\v{x}, t) = e^{i H_0 t} s_e^i(\v{x}, 0) e^{- i H_0 t}$, where $H_0$ is the full electronic Hamiltonian ($H_0 \ket{e} = E_e \ket{e}$), the commutator in the integral in \Eq{kubo_1} simplifies to
\begin{align}
\sum_e p_e \langle e | \, [s_e^i(\v{x}, t), s_e^j(\v{y}, 0)] \, | e\rangle &= \sum_{e e'} p_e \left(e^{i(E_e - E_{e'}) t} \bra{e} s_e^i(\v{x}, 0) \ket{e'} \bra{e'} s_e^j(\v{y}, 0) \ket{e} - (e \leftrightarrow e') \right) \nonumber \\
&= \sum_{e e'} (p_e - p_{e'}) \, e^{i(E_e - E_{e'}) t} \bra{e} s_e^i(\v{x}, 0) \ket{e'} \bra{e'} s_e^j(\v{y}, 0) \ket{e}
\end{align}
where we switched the labels of $e$ and $e'$ in the second step. Performing the time integral in \Eq{kubo_1} with an $i \epsilon$ regulator and then the space integral yields 
\begin{equation} \label{eq:sus_expr}
\x_{ij}(\qv, \w) = - \frac{1}{V} \sum_{e, e'} (p_e - p_{e'}) \, \frac{\bra{e} \tilde{s}_e^i(\qv) \ket{e'} \bra{e'} \tilde{s}_e^j(-\qv) \ket{e}}{\w + E_e - E_{e'} + i \epsilon} \, ,
\end{equation}
where $\tilde{s}_e^i(\mathbf{q}) \equiv \int \dd^3 \mathbf{x} \, e^{-i \mathbf{q} \cdot \mathbf{x}} \, s_e^i(\mathbf{x}, 0)$. By the Sokhotski--Plemelj formula, the anti-Hermitian part of the susceptibility is 
\begin{equation} \label{eq:antiherm_part}
\x_{ij}''(\qv, \w) = \frac{\pi}{V} \sum_{e, e'} (p_e - p_{e'}) \, \bra{e} \tilde{s}_e^i(\qv) \ket{e'} \bra{e'} \tilde{s}_e^j(-\qv) \ket{e} \, \delta\left(\w - (E_{e'} - E_e ) \right).
\end{equation}
In a DM detection experiment, the sample is cooled to negligible temperature, so that $p_e \approx 1 \, (0)$ for filled (unfilled) electronic states. In this case, \Eq{antiherm_part} simplifies to \Eq{chi_pp} of the main text.

\subsection{Models of Spin Response}
\label{app:spin_resp_model}

The absorptive part $\x_{ij}''$ of the (spin) magnetic susceptibility includes a sum over all possible final states. For example, in a magnetically ordered medium, $\ket{e'}$ could be a one-magnon state, describing the excitation of a single magnon. However, it could also be the ground state, describing elastic scattering, or an arbitrary multi-magnon state. If the medium is instead paramagnetic, $\x_{ij}''$ contains contributions from paramagnons, and for a conducting medium it contains contributions from Stoner excitations. In principle, $\x_{ij}''$ even contains contributions from non-magnetic excitations. For instance, phonons can be excited through a coupling to spins in ``magnetovibrational'' scattering.

For concreteness, we focus on insulating, magnetically ordered media with localized electrons, in which case magnon production dominates. If each lattice site $\v{x}_j$ is associated with a spin $\v{S}_j$ due to the spin of the electrons, then in the dipole approximation with quenched orbital angular momentum~\cite{boothroyd2020},
\be
\mathbf{s}_e(\qv) = \sum_j e^{-W_j(\qv)} \, f_j(\qv) \, e^{-i \qv \cdot \v{x}_j} \, (2 \, \v{S}_j)
~,
\ee
where $W_j(\qv)$ is the Debye--Waller factor, accounting for the motion of the lattice ion, and $f_j(\qv)$ is the magnetic form factor, accounting for the electron's spatial distribution about the ion.

As a concrete example, consider a cubic Heisenberg ferromagnet where each spin has magnitude $S$, and all spins are aligned along the $z$-axis. For simplicity, we assume $q$ is low enough to neglect suppressions due to the Debye--Waller factor and the magnetic form factor. (However, for DM masses $m_\x \gtrsim \MeV$, the magnetic form factor can entail a significant suppression, as was discussed in Ref.~\cite{Marocco:2025eqw}.) Then for single-magnon excitation, one can show~\cite{boothroyd2020} that within linear spin wave theory, the only nonzero elements of $\x''_{ij}$ are those given in \Eq{one_magnon_1} of the main text. (In these equations, the spin density $n_s$ is equal to $S/\Omega_c$, where $\Omega_c$ is the volume of the unit cell.) Substituting this into \Eq{scattering_rate} recovers the scattering rate found for the ``$n = 1$'' model in Ref.~\cite{Trickle:2019ovy}.

More generally, YIG is well-described by the Heisenberg model, but has $N = 20$ independent spins per magnetic unit cell, and hence $N$ magnon branches. The toy model above describes excitation of a magnon in the ungapped branch. More generally, $\x''_{ij}$ should include a sum over the $N$ magnon branches, with momentum-dependent magnon polarization factors. Accounting for this would recover the ``full model'' results previously found in Ref.~\cite{Trickle:2019ovy}.

This model only accounts for single-magnon excitation, because other processes are subdominant or irrelevant. For example, particles can undergo elastic scattering off the magnet's static magnetization, which yields a contribution to $\x_{zz}''$. However, such events would not be detectable in DM experiments because they deposit no energy. In addition, particles can gain energy through magnon absorption, but in the experiment of interest~\cite{princep2017full}, the YIG sample was cooled to $T \lesssim 10 \, \mathrm{K}$ so that thermal excitation was negligible, and this would be the case for any DM detector as well. 

It is also interesting to consider two-magnon excitation, first discussed in the context of DM scattering by Ref.~\cite{Esposito:2022bnu}. For small momentum transfers $qa \ll 1$, where $a$ is the lattice spacing, the two-magnon excitation rate is suppressed by powers of $qa$. However, in the very light regime $m_\x \sim \keV$ this process can dominate over single magnon excitation in antiferromagnets, for kinematic reasons.

By contrast, for the neutron scattering experiments of interest here, we generally operate in the regime $qa \gtrsim 1$, as shown in \Fig{data_kinematics_summary}, where there are no such kinematic restrictions. In this case, two-magnon excitation can be observed~\cite{tennant2003neutron,huberman2005two} but is subdominant. As discussed in Refs.~\cite{huberman2005two,chan2023neutron}, this follows very generally from sum rules, which state that the multi-magnon excitation rate is parametrically suppressed, relative to the single-magnon excitation rate, by $\sim \Delta S / S$, where $\Delta S$ is the longitudinal spin fluctuation. The magnets considered in those two works have $S = 5/2$, and have $\Delta S \sim 0.2$, enhanced by two-dimensional and helicoidal structure respectively. For YIG, we also have $S = 5/2$, and expect an even smaller $\Delta S$ because it is three-dimensional with collinear order, so multi-magnon excitation is suppressed by more than an order of magnitude. Nonetheless, a strength of our data driven approach is that it can be used to infer the DM scattering rate even when such processes are important. 

\section{Dominance of Coupling to Electron Spin}
\label{app:subdominant_ints}

Here we review why the dominant interaction for neutron scattering is the neutron-electron spin-spin interaction, and how this extends to the DM models we consider. 

First, the neutron has other electromagnetic interactions. Its magnetic dipole moment also couples to nuclei; however, the magnetic dipole moment of a nucleus is suppressed, relative to that of an electron, by $\sim m_e / m_p \sim 10^{-3}$. in addition, the neutron has other electromagnetic moments, such as a polarizability and charge radius. However, their effects are all small due to the neutron's small size, and correct the scattering rate by $\sim 10^{-3}$~\cite{sears1986electromagnetic}. 

Second, the neutron also has a direct interaction with atomic nuclei. For a neutron scattering on a single atom, these effects are roughly comparable in strength, since the nuclear interaction depends on the nuclear length scale $r_N \sim \text{fm}$, while the electromagnetic interaction depends on the classical electron radius $\alpha_e / m_e \sim \text{fm}$. However, in a material, the nuclear interaction primarily excites phonons, with a rate parametrically suppressed by $q^2 / (m_N \omega)$ where $m_N$ is the nuclear mass~\cite{Trickle:2019nya}. For benchmark values $m_N \sim 50 \, \mathrm{GeV}$ and $\omega \sim 20 \, \mathrm{meV}$, this corresponds to a factor of $(q / 30 \, \keV)^2$, which is a strong penalty for all $q$ in the dataset. (We note, however, that at the highest $q$ covered, magnetic scattering can be suppressed to an even greater extent by the magnetic form factor.)

We therefore focus on the neutron's interaction with electrons through its magnetic dipole moment, shown in \Eq{L_ne}. To quickly extract the terms in the nonrelativistic interaction Hamiltonian, we use Table I of Ref.~\cite{Trickle:2020oki}, which gives the nonrelativistic limits of the magnetic dipole and vector currents. We find that for a nonrelativistic neutron and electron with initial velocities $\mathbf{v}_i$ and spin operators $\mathbf{S}_i$, 
\begin{equation}
\mathcal{H}_{ne}^{\text{eff}} \propto J^\mu_{\text{mdm}}(\qv) \, \frac{1}{|\qv|^2} J_{\mu, V}(-\qv) \propto \dfrac{(\hat{\qv} \times \mathbf{S}_n) \cdot (\hat{\qv} \times \mathbf{S}_e)}{m_e \, m_n} + \frac{1}{4 \, m_n^2} + \frac{\mathbf{S}_n \cdot (i \hat{\qv} \times \vv_e)}{|\qv| \, m_n} - \frac{\mathbf{S}_n \cdot (i \hat{\qv} \times \vv_n)}{|\qv| \, m_n}.
\end{equation}
This result is compatible with Table II of Ref.~\cite{Trickle:2020oki}, up to a sign error on that work's coefficient $c_{5b}^{(\psi)}$, and also matches known results in the neutron scattering literature~\cite{sears1986electromagnetic,boothroyd2020}.

The first term is the spin-spin coupling we kept in Sec.~\ref{subsec:ns_derivation}. The second term is a contact interaction called the Foldy interaction, which is suppressed by $\sim m_e / m_n$ and hence negligible. The third term is the coupling of the neutron's magnetic moment to the electron's orbital magnetization, which in some cases can be comparable to the electron's spin magnetization; however, orbital magnetization is quenched in rare earth compounds such as YIG. The final term is the spin-orbit interaction. For an order-one momentum transfer $|\qv| \sim m_n v_n$, it is suppressed by $\sim m_e / m_n$ and hence negligible. Equivalently, it is only relevant for very small momentum transfers $|\qv| \sim m_e v_n \sim 0.01 \, \mathrm{keV}$, which are well below the resolution of the neutron scattering dataset. Thus, the spin-spin interaction in dominates for the kinematic regimes and materials of interest here. 

Finally, let us explain why the DM models in Table~\ref{tab:models} dominantly couple to electron spin. For the bottom four rows, the DM coupling to electron spin is the leading interaction in the nonrelativistic limit, with other interactions suppressed by powers of $v_\x$ or $v_e$. For the magnetic dipole DM and anapole DM models, additional terms appear at the same order in velocities~\cite{Catena:2019gfa,Trickle:2020oki}. These terms do not directly couple to the electron spin, and would predominantly excite phonons. However, we have seen that the phonon production rate is suppressed by $q^2/(m_N \omega)$, which is a strong penalty for $m_\x \lesssim \text{MeV}$. On the other hand, if $m_\x \gg \text{MeV}$, then these terms are suppressed by the small ratio $m_e/m_\x$. Thus, in either case the dominant process is magnon production through the DM's coupling to electron spins. 

\section{The Neutron Scattering Dataset}
\label{app:data}

\begin{figure}
\centering
\includegraphics[width=0.49\linewidth]{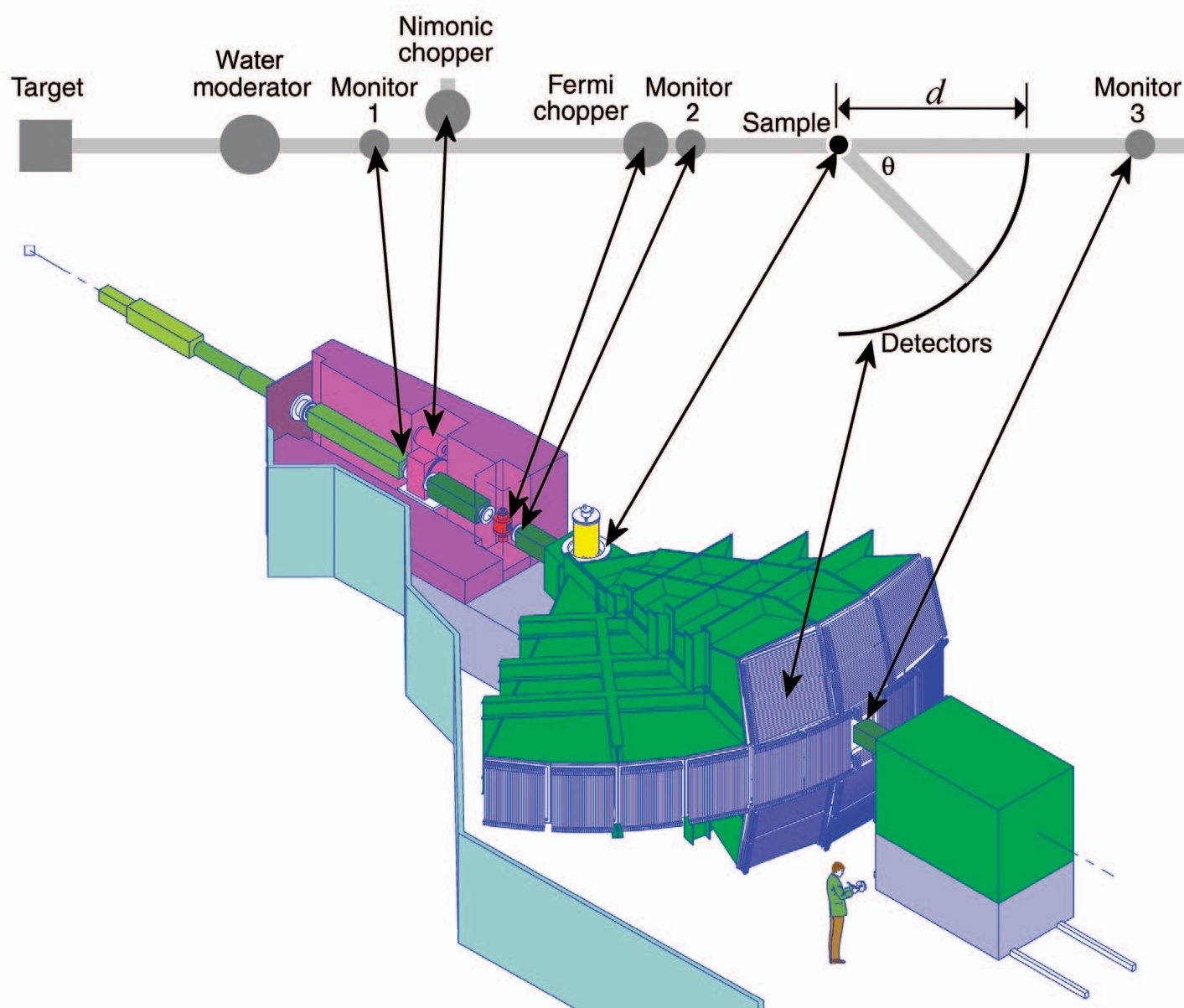}
\raisebox{-3mm}{\includegraphics[width=0.49\linewidth]{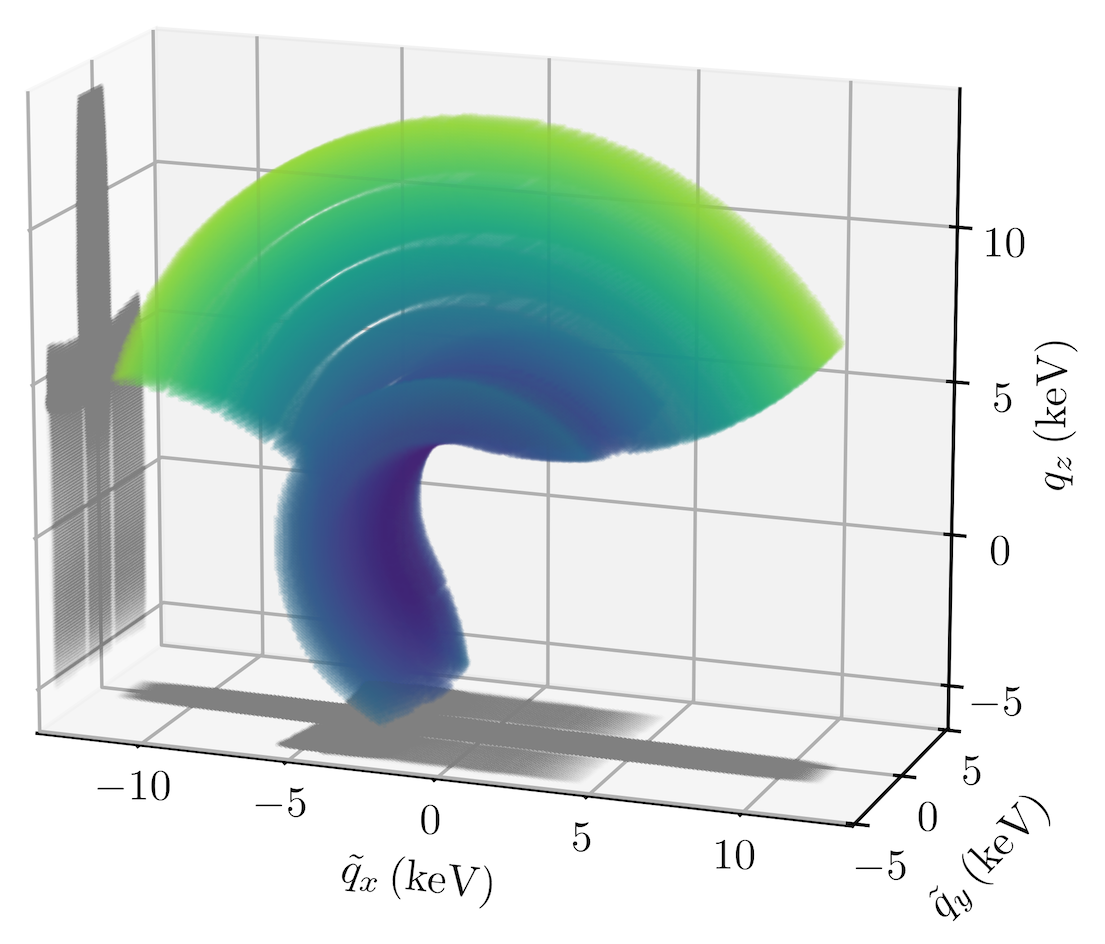}}
\caption{\textbf{Left:} Schematic of the MAPS spectrometer, reproduced from Ref.~\cite{maps_schematic}. \textbf{Right:} The momentum transfers $\mathbf{q}$ probed by the neutron scattering data, in the crystal frame, for a fixed energy transfer $\omega = 50 \, \mathrm{meV}$. To aid visualization, the color indicates the magnitude of $\mathbf{q}$, and two projections of the data are shown. As discussed in the text, 
the shape of this region can be understood from the shape of the detector and the way the crystal was rotated during measurement.}
\label{fig:3d_map}
\end{figure}

\begin{figure}
\centering
\includegraphics[width=0.6\linewidth]{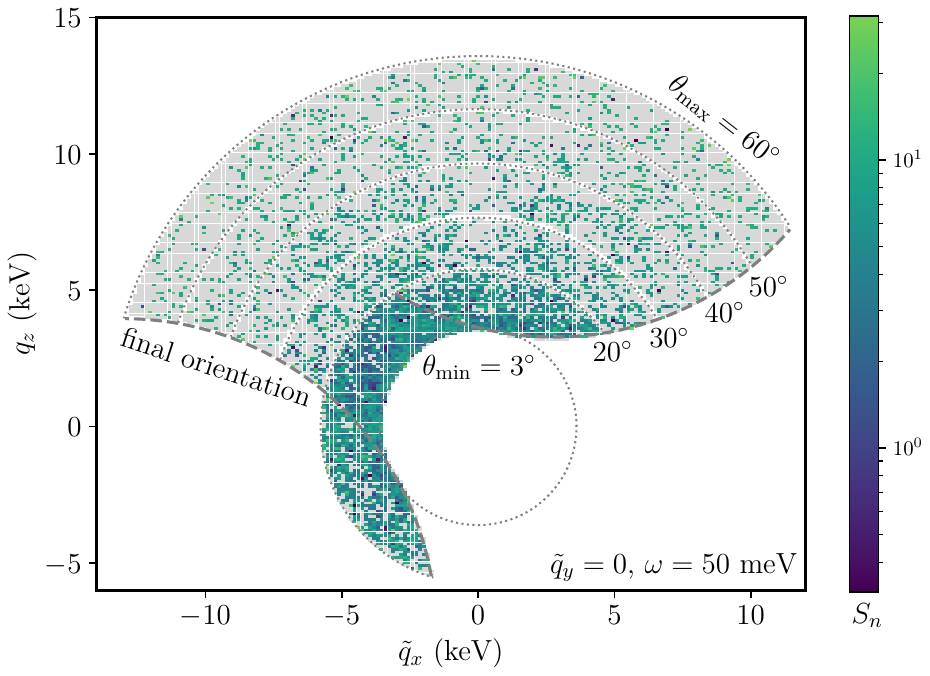}
\caption{A slice of the neutron scattering dataset at fixed $\tilde{q}_y$ and $\omega$. Dotted circles indicate the momentum transfers corresponding to a fixed angular deflection $\theta$. Dashed curves show the momentum transfers probed by a given crystal orientation; the rotation of the crystal rotated this curve in the plane of the page. Bins are shaded according to their inferred dynamic structure factor $S_n$ (arbitrary units). Grey bins were covered by the dataset but contained zero counts.}
\label{fig:data_slice}
\end{figure}

Here we elaborate on the discussion in Sec.~\ref{subsec:connection_to_data}, by providing some visualizations of the neutron scattering dataset of Ref.~\cite{princep2017full}. The MAPS detector is shown in blue in the left-panel of \Fig{3d_map}. As we have already noted, small deflection angles $\theta < 3^\circ$ are not covered, as the beam exits through this direction. For $3^\circ < \theta < 20^\circ$, a low angle detector bank provides relatively complete angular coverage. In addition, there is a high angle detector bank extending through the range $20^\circ < \theta < 60^\circ$, though it covers only a relatively narrow horizontal strip. 

We show the three-dimensional $\qv$ covered by the dataset, at the representative energy transfer $\omega = 50 \, \mathrm{meV}$, in the right-panel of \Fig{3d_map}. For clarity we rotate the coordinate system, relative to that directly used by the software, by defining $\tilde{q}_x = (q_x + q_y)/\sqrt{2}$ and $\tilde{q}_y = (q_x - q_y)/\sqrt{2}$. The figure shows that for larger $|\qv|$, the range of $\tilde{q}_y$ values covered is relatively small. This is because in the laboratory frame, the crystal was always oriented so that $\tilde{q}_y$ was approximately the vertical component of the momentum transfer, and a scattered neutron with large vertical momentum transfer would miss the high angle bank's detector strip. To cover a wider range of $\tilde{q}_y$, one could use another instrument with more detector coverage, or rotate the crystal about other axes. 

Next, we examine a slice of the dataset at fixed $\tilde{q}_y = 0$, as shown in \Fig{data_slice}. The range of $|\qv|$ covered by the data is consistent with the discussion above. In addition, there are gaps in the dataset at deflection angles of $20^\circ$, $30^\circ$, $40^\circ$, and $50^\circ$, which correspond to the gaps between detector panels visible in the left-panel of \Fig{3d_map}. For each crystal orientation, the momenta probed form an arc, which is asymmetric because the wide angle detector bank only covers one side. As the crystal is rotated, this arc is rotated in the $(\tilde{q}_x, q_z)$ plane through a total range of $\sim 120^\circ$, matching the discussion in Ref.~\cite{princep2017full}. Rotating the crystal through a full $360^\circ$ would extend data coverage to the full annular region $3^\circ < \theta < 60^\circ$.

At larger $|\qv|$, many of the bins contain zero counts. To understand this, we note that the bins are cubical in momentum space with side length $0.1 \, \mathrm{keV}$, roughly corresponding to $\sim 1^\circ$ at the highest momentum transfers. (The bins appear as rectangular in \Fig{data_slice} simply because it is plotted in a rotated coordinate system.) However, the crystal was rotated through steps of $0.25^\circ$~\cite{princep2017full} and the angle subtended by a detector element is $(2.5 \, \mathrm{cm}) / (6 \, \mathrm{m}) \sim 0.2^\circ$~\cite{MAPS}. These are both much smaller than the bin width, so the zero counts are not an artifact of binning too finely. Instead, they reflect a lack of statistics: at large $|\qv|$, there are many bins for a given $|\qv|$, each one is sampled for a shorter time as the crystal is rotated, and the scattering rate itself is penalized by falling magnetic form factors.

The large number of bins with zero counts makes it difficult to distinguish fine features at large $|\qv|$, but it does not bias our results, since we include these bins when we construct the angular average $\bar{S}_n(q, \omega)$ in Sec.~\ref{subsec:sensitivity_proj}. From \Fig{data_slice}, one can see that the data is roughly isotropic at large $|\qv|$, justifying our angular extrapolation procedure. In addition, the number of bins with nonzero counts is still extremely large ($\gtrsim 10^7$), so that statistical uncertainties in our final results are small. To reduce the number of zero counts, one can measure for a longer time or increase the bin size. 

\end{document}